\documentclass{aa}

\usepackage{txfonts}
\usepackage{graphics}
\usepackage{epsfig}

\bibpunct{(}{)}{;}{a}{}{,}

\begin{document}

\title{The VMC survey - IX. Pilot study of the proper motion of stellar populations in the LMC from 2MASS and VISTA data\thanks{Based on observations made with VISTA at the Paranal Observatory under program ID 179.B-2003.}}

\author{M.-R.L. Cioni\inst{1, 2, 3}\thanks{Research Fellow of the Alexander von Humboldt Foundation}
	\and L. Girardi\inst{4}
	\and M.I. Moretti\inst{5}
	\and T. Piffl\inst{1}
	\and V. Ripepi\inst{5}
	\and S. Rubele\inst{4}	
	\and R.-D. Scholz\inst{1}	
	\and K. Bekki\inst{6}
	\and G. Clementini\inst{7}
	\and V.D. Ivanov\inst{8}
	\and J.M. Oliveira\inst{9}
	\and J.Th. van Loon\inst{9}
}

\offprints{mcioni@aip.de}

\institute{
	Leibnitz-Institut f\"{u}r Astrophysik Potsdam, An der Sternwarte 16, 14482 Potsdam, Germany
	\and University Observatory Munich, Scheinerstrasse 1, 81679 Munich, Germany
	\and University of Hertfordshire, Physics Astronomy and Mathematics, Hatfield AL10 9AB, United Kingdom
	\and INAF, Osservatorio Astronomico di Padova, vicolo dell'Osservatorio 5, 35122 Padova, Italy
	\and INAF, Osservatorio Astronomico di Capodimonte, via Moiariello 16, 80131 Napoli, Italy
	\and ICRAR, M468, University of Western Australia, 35 Stirling Hwy, Crawley 6009, Western Australia
	\and INAF, Osservatorio Astronomico di Bologna, via Ranzani 1, 40127 Bologna, Italy
	\and European Southern Observatory, Av. Alonso de C\'{o}rdoba 3107, Casilla 19, Santiago, Chile
	\and Lennard-Jones Laboratories, Keele University, School of Physical and Geographical Science, ST5 5BG, United Kingdom
	}

\date{Received 18 June 2013 / Accepted 28 October 2013}

\titlerunning{SEP proper motion}

\authorrunning{Cioni et al.}

\abstract{Proper motion studies are fundamental ingredients in the understanding of the orbital history of galaxies. Current measurements do not yet provide a satisfactory answer to the possible scenarios for the formation and evolution of the Magellanic Clouds and of the Bridge and Stream that link them with each other and our Galaxy.}{We use multi-epoch near-infrared observations from the VISTA survey of the Magellanic Cloud system (VMC) to measure the proper motion of stars of the Large Magellanic Cloud (LMC), in one tile of $1.5$ deg$^2$ centred at $(\alpha, \delta) = (05$:$59$:$23.136$, $-66$:$20$:$28.68)$ and including the South Ecliptic Pole, with respect to their 2MASS position over a time baseline of $\sim 10$ years. Proper motions from VMC observations only, spanning a time range of $\sim1$ year, are also derived.}{Stars of different ages are selected from the colour-magnitude diagram, $(J-K_\mathrm{s})$ vs. $K_\mathrm{s}$, and their average coordinate displacement is computed from the difference between $K_\mathrm{s}$ band observations from VMC and 2MASS or among VMC data alone for stars as faint as $K_\mathrm{s}=19$ mag. Proper motions are derived by averaging up to seven 2MASS-VMC combinations in the first case and from the slope of the best fit line among the seven VMC epochs in the second case. Separate proper motion values are obtained for Cepheids, RR Lyrae stars, Long Period Variables and Eclipsing Binary stars in the field.}{The proper motion of $\sim40,000$ LMC stars in the tile, with respect to $\sim 8000$ background galaxies, obtained from VMC data alone, is $\mu_\alpha cos(\delta) = +2.20\pm0.06$ (stat) $\pm0.29$ (sys) and $\mu_\delta = +1.70\pm0.06$ (stat) $\pm0.30$ (sys) mas yr$^{-1}$. This value agrees with recent ground-based determinations but is larger than studies with the HST; the cause of this discrepancy may be due to additional systematic errors in the data. Our result implies either higher tangential motion or higher internal motion or the combination of these, though we can not discuss these possibilities based on one field quantitatively. The proper motion of the LMC is also clearly distinct from the proper motion derived for stars of the Milky Way foreground. The relative proper motion between the foreground stars and the LMC stars is $\sim5$ mas yr$^{-1}$.  Furthermore, we measure a decrease of the proper motion with increasing logarithm of stellar age for LMC stars.}{This study, based on just one VMC tile, shows the potential of the 2MASS-VMC and VMC-VMC combinations for a comprehensive investigation of the proper motion across the Magellanic system.}

\keywords{Surveys - Magellanic Clouds - Infrared: stars - Proper motions}

\maketitle

\section{Introduction}

The Large Magellanic Cloud (LMC) and the Small Magellanic Cloud (SMC) are neighbouring dwarf irregular galaxies to the Milky Way (MW) and represent a close-by ($50-60$ kpc) laboratory for studies of stellar evolution and galaxy interaction. Their history couples them with each other and with the MW. Dynamical interactions  are responsible for various episodes of star formation and most likely also for the formation and/or shaping of the bars \citep[e.g.][]{2012MNRAS.421.2109B} as well as of other sub-structures, e.g. a ring-like feature surrounding the SMC \citep{2004AJ....127.1531H, 2006A&A...448...77C}. The detection of accreted SMC stars onto the LMC \citep{2011ApJ...737...29O} supports these processes. The link between dynamical processes (orbits, internal kinematics and environmental effects due to the MW and the Local Group), star forming events and structure (disk, spheroid, halo) as well as sub-structure formation (streams, bars and other tidal features) is not yet properly established and represents a major uncertainty in the understanding of the formation and evolution of the Magellanic Clouds. The formation mechanism of the two major tidal features: a Bridge, connecting the two galaxies traced by neutral hydrogen, but with stars associated to it \citep{1985Natur.318..160I, 2013A&A...551A..78B, 2013ApJ...768..109N} and a Stream, to-date still purely gaseous and without an identified stellar component, sweeping $\sim 200^\circ$ across the southern sky and probably emanating from the SMC \citep{2010ApJ...723.1618N}, is still unclear; one example of a successful interpretation of the formation and shape of the Stream would imply two pericentre passages by the Milky Way \citep{2012MNRAS.422.1957B}.

Proper motion measurements with the {\it Hubble Space Telescope} (HST) indicate that the Magellanic Clouds may have entered the MW potential only recently \citep[e.g.][]{2013ApJ...764..161K}. A first infall scenario implies that the Magellanic Clouds are bound to each other, the SMC is on an elliptical orbit around the LMC, the LMC has a relatively large mass ($>10^{11}$ M$_\odot$) and the MW mass is small ($<1.5\times 10^{12}$ M$_\odot$). In this framework, the Bridge and in particular the Stream, would result from the interaction between the Magellanic Clouds \citep[e.g.][]{2010ApJ...721L..97B} rather than with the MW, contrary to expectations \citep[e.g.][]{2012ApJ...750...36D}. The HST results, which refer to a mixed stellar population (with $\sim30$ stars per field) and sample a small fraction of the Magellanic Clouds (about two dozen fields of $0.25$ arcmin$^2$ each in size), may be affected by local random motion. The small size of the HST sample may bias the proper motion towards the motion of blue (or red) stars, that may (or may not)  move accordingly to their parent gas cloud and under the influence of their host structure. Furthermore, the LMC has a thick disk with a bar \citep{2002AJ....124.2639V} and stars that deviate from circular motion.  \citet{2011ApJ...730L...2B} argued that $>300$ homogeneously distributed measurements of the local kinematics are needed to derive accurate averaged centre-of-mass proper motions and our comprehensive investigation following this pilot study (see below), contrary to the HST one, overcomes this limitation. HST proper motions are also dominated by uncertainties on the internal kinematics and structure of the Magellanic Clouds \citep{2013arXiv1305.4641V} and the derived orbits depend strongly on the MW and LMC masses that are still uncertain. On the contrary, large statistical samples from ground based observations have big uncertainties in the measured motions and are, at present, restricted to the outer regions of the Magellanic Clouds \citep[e.g.][]{2010AJ....140.1934V, 2011AJ....141..136C} or limited to a specific type of stars, i.e. oxygen-rich (O-rich) asymptotic giant branch (AGB) stars in \citep{2010AJ....140.1934V}.

In this study the proper motion of the LMC is measured from the combination of 2MASS and VISTA near-infrared data that span a time range of $\sim10$ years as well as from VISTA data alone across a time baseline of $\sim1$ year. The data are of sufficient quality to provide at the same time a large statistical sample of targets and a high accuracy in the measured motions. This is the first of a series of studies on the proper motion of the Magellanic Clouds, using data from the VISTA survey of the Magellanic Clouds system \citep[VMC;][]{2011A&A...527A.116C}, that aim to put firm constraints on the internal kinematics of the galaxies, the distribution of mass, their mutual interaction and the formation of tidal features. This pilot study is focused on one out of $68$ VMC tiles of $\sim 1.5$ deg$^2$ each in size covering the LMC and the large stellar sample available argues in favour of obtaining reliable kinematics also in sub-sets of tiles. In Sect.~\ref{data} we describe the 2MASS and VMC epochs of observations in the $K_\mathrm{s}$ band and the sample of stars used to measure the VMC-2MASS proper motion in a North-East field of the LMC disk. Section \ref{pm} shows the proper motion values for stars of a different type selected from the colour-magnitude diagram, $(J-K_\mathrm{s})$ vs. $K_\mathrm{s}$, accordingly to the population boxes defined in \citet{2000ApJ...542..804N} for the 2MASS stars. Section~\ref{vpm} shows instead the proper motion values for fainter VMC stars selected from the same diagram accordingly to new boxes established from the analysis of the star formation history by \citet{2012A&A...537A.106R}. The proper motion of samples of known variable stars is also derived and the proper motion trend with stellar ages is analysed. Section \ref{discussion} compares the results with previous measurements in the LMC and the predictions for the foreground stars obtained from the GALAXIA model of the MW. Conclusions and future work are outlined in Sect.~\ref{conclusion}.

\section{Data}
\label{data}

We focus on one region, tile LMC $8\_8$ from the VMC survey centred at $(\alpha, \delta) = (05$:$59$:$23.136, -66$:$20$:$28.68)$, located in the outer disk of the LMC and including the South Ecliptic Pole. This tile was chosen because it was the first tile fully observed by the VMC survey and it is characterised by a low line-of-sight reddening and crowding. For a description of the VMC tile pattern covering the Magellanic system see \citet{2011A&A...527A.116C}.

\subsection{2MASS data}
\begin{figure}
\resizebox{\hsize}{!}{\includegraphics{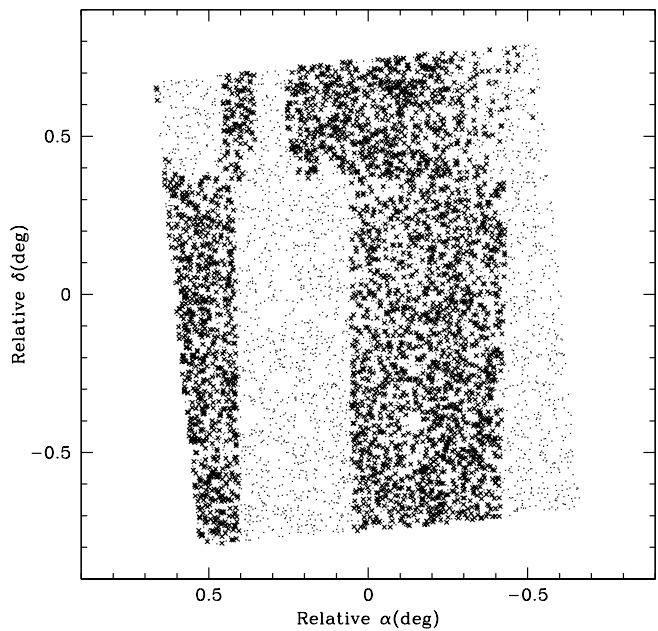}}
\caption{Surface distribution of stars from 2MASS All-sky (points) and $6\times$ (crosses) catalogues for tile LMC $8\_8$. The $6\times$ catalogue refers to sources with $K_\mathrm{s}>14.3$ mag.} The centre of the tile is at $(\alpha_0$, $\delta_0)=(-89^\circ.8464$, $-66^\circ.3413)$.
\label{map88}
\end{figure}

The 2MASS All-Sky and the 2MASS Long Exposure ($6\times$) Scan Databases \citep{2006AJ....131.1163S} were used to select counterparts for the VMC sources.
In the 2MASS All-Sky catalogue only sources with S/N$>$$10$ and of high quality were selected. These correspond to sources with $J<15.8$ mag, $H<15.1$ mag and $K_\mathrm{s}<14.3$ mag and quality flags {\it rd\_flg} $=222$ (magnitudes mesured using profile fitting photometry performed simultaneously on six individual $1.3$ s exposures covering the sources), {\it ph\_qual} $=\mathrm{AAA}$ (valid detections within the range of magnitudes, photometric uncertainties and rd$\_\mathrm{flg}$ values above), {\it bl\_flg} $=111$ (single-profile fits to isolated sources) and {\it cc\_flg} $=000$ (detected sources unaffected by artefacts). The 2MASS $6\times$ catalogue was then used to extract sources that are on average a magnitude fainter than the main survey and that obey the same quality criteria for $K_\mathrm{s}>14.3$ mag. Their surface distribution, however, is not as homogeneous as that of the main survey (Fig. \ref{map88}). The single-epoch 2MASS observations have at best a PSF size of $2.5^{\prime\prime}$ and those used in this study were obtained from  October 1998 to February 2000 (All-Sky) and from December 2000 to February 2001 ($6\times$). Each source was imaged at least six times following a sub-pixel dithering technique, over a pixel size of $2^{\prime\prime}$, that improved the spatial resolution of the final coadded image.
The astrometry of the All-Sky catalogue in the International Celestial Reference System was evaluated by comparing 2MASS sources with those in the Tycho-$2$ \citep{2000A&A...355L..27H} and UCACr$10$ catalogues\footnote{For more details on the 2MASS All-sky astrometry see http://spider.ipac.caltech.edu/staff/hlm/2mass/overv/overv.html.}.  For sources with {\it rd\_flg} $=222$ the positional accuracy, reflecting both systematic and random errors, is $70-80$ mas with respect to these catalogues; \citet{2005A&A...429..739D} derived a standard deviation of $90$ mas. The astrometry of the $6\times$ catalogue was estimated with reference to the All-Sky catalogue with respect to the UCAC$2$ \citep{2004AJ....127.3043Z} catalogue and has a mean radial offset of $95$ mas similar to the 2MASS main survey\footnote{For more details on the $6\times$ 2MASS astrometry see http://www.ipac.caltech.edu/2mass/releases/allsky/doc/seca3\_2c.html.}.\\

\subsection{VMC data}
\label{vmcdata}
The VMC data analysed in this study refers to observations acquired with the Visible and Infrared Survey Telescope for Astronomy \citep[VISTA;][]{2010Msngr.139....2E} from November $2009$ to November $2010$. The data were reduced onto the VISTA photometric system, which is close to a Vegamag one, with the VISTA Data Flow System pipeline v1.1 \citep[VDFS;][]{2004SPIE.5493..411I} and extracted from the VISTA Science Archive \citep[VSA;][]{2012A&A...548A.119C}. The VMC survey strategy involves repeated observations of tiles across the Magellanic system, where one tile covers uniformly an area of $\sim1.5$ deg$^2$ in a given wave band with $3$ epochs at $Y$ and $J$, and $12$ epochs at $K_\mathrm{s}$ spread over a time range of a year or longer. Eleven of the $K_\mathrm{s}$ epochs refer to the monitoring campaign and correspond each to exposure times of $750$ s (deep) while the other corresponds to two observations with half the exposure time (shallow) that are not necessarily obtained during the same night. Additional observations may be present, for example to recover observations redone when original ones did not meet the quality requirements. Details about the observing strategy and the data reduction are given in \citet{2011AJ....141..136C}.

Sources were first selected from the ``vmcsource" merged catalogue containing sources extracted from "tiledeepstacks" in the $Y$, $J$ and $K_\mathrm{s}$ bands, respectively. 
These are deep tile images resulting from the combination of individual tile images taken at different observing epochs. Only objects detected in all three wave bands were considered. These correspond to $61$\% of the total. Each tile is the result of stacking six individual ``pawprint" observations, each containing $16$ detector images obtained from the combination of multiple exposures at different jitter positions \citep{2004SPIE.5493..411I}. This process uses a ``dribbling" technique to distribute the information of overlapping pixels onto the grid of the tile \citep{2009UKIRT.25.15}. The astrometric distortions in VISTA images are corrected based on 2MASS data and the residuals are hence dominated by 2MASS errors. On the other hand, we can safely assume that the VISTA and 2MASS data are on the same system.
The astrometry of tiles suffers from a $10-20$ mas systematic pattern due to residual World-Coordinate-System errors from the pawprints and a residual radial distortion of up to $\pm100$ mas across the field due to an inconsistent use of the Zenithal-Polynomial projection (this software bug has been corrected in further processings). Subsequently, the same sources were extracted from the multiple individual tile images, ``tilestacks", in the deep $K_\mathrm{s}$ observations and only those with photometric uncertainties $<0.1$ mag in $K_\mathrm{s}$ at each epoch were retained. The VMC data were obtained under homogeneous sky conditions since they are acquired in service mode when the sky quality meets the requested VMC criteria, see \citet{2011A&A...527A.116C}. The average tile quality of the VMC $K_\mathrm{s}$-band data analysed here corresponds to: $0.34^{\prime\prime}$ pixel size, $0.91^{\prime\prime}$ FWHM, $0.05^{\prime\prime}$ ellipticity, $1.38\pm0.05$ airmass and a magnitude sensitivity of $18.77$ for sources with photometric errors $<0.1$ mag.

\subsection{VMC-2MASS sample selection}

\begin{figure}
\resizebox{\hsize}{!}{\includegraphics{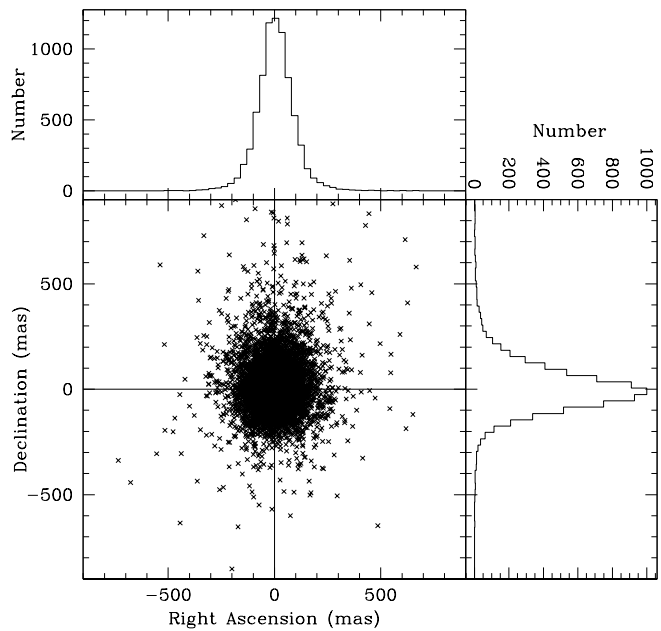}}
\caption{Distribution of the coordinate differences between positionally matched VMC and 2MASS sources in tile LMC $8\_8$. Histograms have bins of $30$ mas in size.}
\label{dist}
\end{figure}

\begin{figure}
\resizebox{\hsize}{!}{\includegraphics{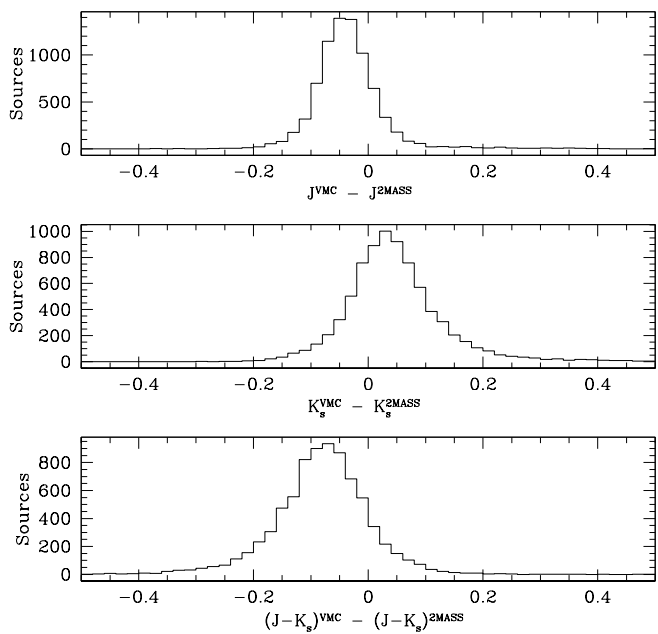}}
\caption{Number distribution of positionally matched VMC and 2MASS sources in tile LMC $8\_8$ as a function of $J$ and $K_\mathrm{s}$ magnitudes as well as $(J-K_\mathrm{s})$ colour. The bin size is of $0.02$ mag.}
\label{diffs}
\end{figure}

\begin{figure}
\resizebox{\hsize}{!}{\includegraphics{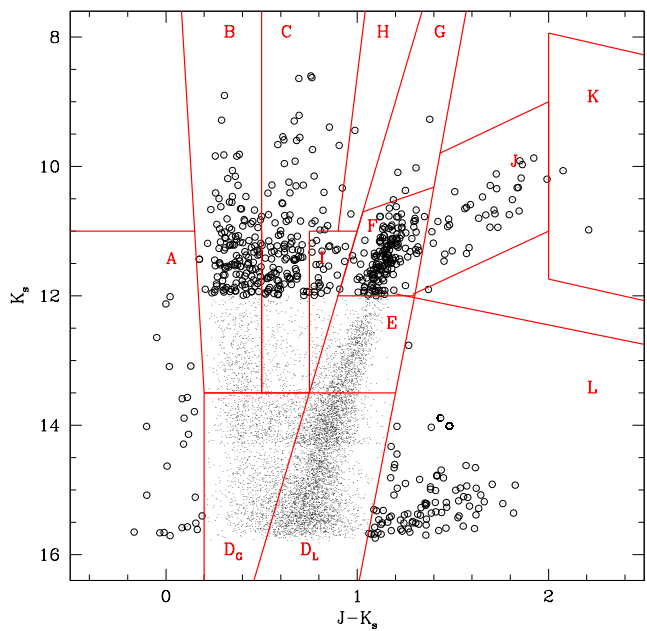}}
\caption{CMD of 2MASS sources positionally matched with the VMC data in tile LMC $8\_8$. Region boundaries according to \citet{2000ApJ...542..804N} and modified as in Table \ref{spop} to distinguish among different types of stars are indicated. Empty circles identify sources in less crowded regions of the diagram and both regions K and L include sources with $(J-K_\mathrm{s})<5$ mag.}
\label{cmd}
\end{figure}

VMC and 2MASS sources were positionally matched within $1^{\prime\prime}$ (Fig. \ref{dist}). There are $7980$ sources in tile LMC $8\_8$ that satisfy the selection criteria above where on average $\sim95$\% are morphologically classified as stellar and $\sim5$\% as extended with a minor contribution of probably stellar objects according to the VDFS pipeline. In order to increase the reliability of the sample we required that differences in magnitudes and colour are within $0.5$ mag. Sources that were excluded have $K_\mathrm{s}<11$ mag and/or $(J-K_\mathrm{s})>1.2$ mag suggesting that it is difficult to assign VMC counterparts automatically to bright stars approaching the VISTA saturation limit ($K_\mathrm{s}\sim10$ mag) and to red extended sources at the faint end of the 2MASS sensitivity (cf. Fig. \ref{cmd}). Bright and red sources are mostly carbon-rich (C-rich) AGB stars which are known to experience variations in magnitude and colour, hence the criteria imposed in this study may be too stringent in securing their cross-correlation. Figure \ref{diffs} shows the histograms of the selected sources. Systematic differences of $\Delta J=-0.05$, $\Delta K_\mathrm{s} = 0.03$ and $\Delta (J-K_\mathrm{s}) = 0.07$ mag are derived between VMC and 2MASS measurements. These systematic differences are not used to transform magnitudes from one system to the other because in the following study 2MASS and VMC magnitudes are used independently.

\begin{figure}
\resizebox{\hsize}{!}{\includegraphics{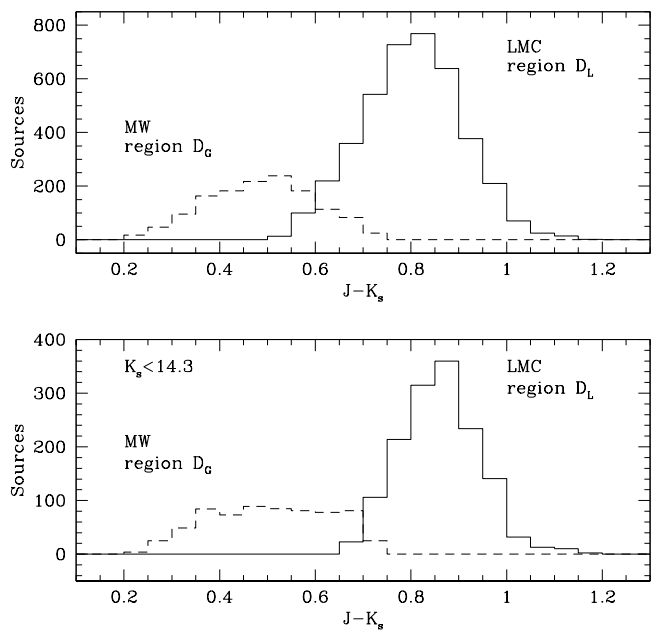}}
\caption{Histograms of sources from VMC tile LMC $8\_8$ contained in region D. The selection criteria outlined in Table \ref{spop} show the two separate distributions of MW and LMC stars for sources from the 2MASS All-Sky catalogue (bottom) and including sources from the 2MASS $6\times$ catalogue (top).}
\label{hist}
\end{figure}

The near-infrared colour-magnitude diagram (CMD), $(J-K_\mathrm{s})$ vs. $K_\mathrm{s}$ in the 2MASS photometric system, of the final sample of $7675$ sources is shown in Fig.  \ref{cmd} . Different types of stars are identified using the criteria from \citet{2000ApJ...542..804N} where the region boundaries listed in their Table $2$ were re-derived from their figures and the results are given in the Appendix. In addition, region D was extended to faint sources and split in two parts to distinguish MW from LMC sources. Figure \ref{hist} shows that LMC red giant branch stars (RGB) stars populating region D follow a symmetric distribution peaked at $(J-K_\mathrm{s})\sim0.8$ mag while MW stars peak at $\sim0.5$ mag. Regions A and L were also extended. 

\section{VMC-2MASS proper motion}
\label{pm}
The proper motion (PM) of each source was derived from the positional difference between VMC and 2MASS coordinates, corresponding to a time range of $\sim10$ yr, such as: $\mu_\alpha cos(\delta) = ((\alpha_\mathrm{VMC}-\alpha_\mathrm{2MASS})\times cos(\delta_\mathrm{VMC}))/\Delta t$ and $\mu_\delta = (\delta_\mathrm{VMC}-\delta_\mathrm{2MASS})/\Delta t$ where $\Delta t = (t_\mathrm{VMC} - t_\mathrm{2MASS})/365.25$ and $t_\mathrm{VMC}$ and $t_\mathrm{2MASS}$ are the Julian days of VMC and 2MASS observations. Up to $7$ independent VMC epochs (see Sect. \ref{vmcvmc}) and one 2MASS epoch are used to derive average PMs for each star. Then, the PM of stars in different CMD regions is obtained from the average of the individual stellar PMs.  An iterative $3\sigma$ clipping technique, with up to $10$ iterations, is applied to reject outliers that may influence the calculation of the averaged values. The final iteration corresponds to the rejection of only a few sources or none, in case of less than $10$ iterations. The resulting mean PMs ($\mu$) are given in Table \ref{tab88}, these are relative PMs not absolute ones (e.g. with respect to non-moving objects). Columns list the region, the number of sources (N) within each region, the mean PM in $\alpha$ and $\delta$ as well as their uncertainty. The uncertainty corresponds to the standard error on the mean $\delta_\mu=\sigma/\sqrt{N}$ where $\sigma$ is the standard deviation.

\begin{figure*}
\resizebox{\hsize}{!}{\includegraphics{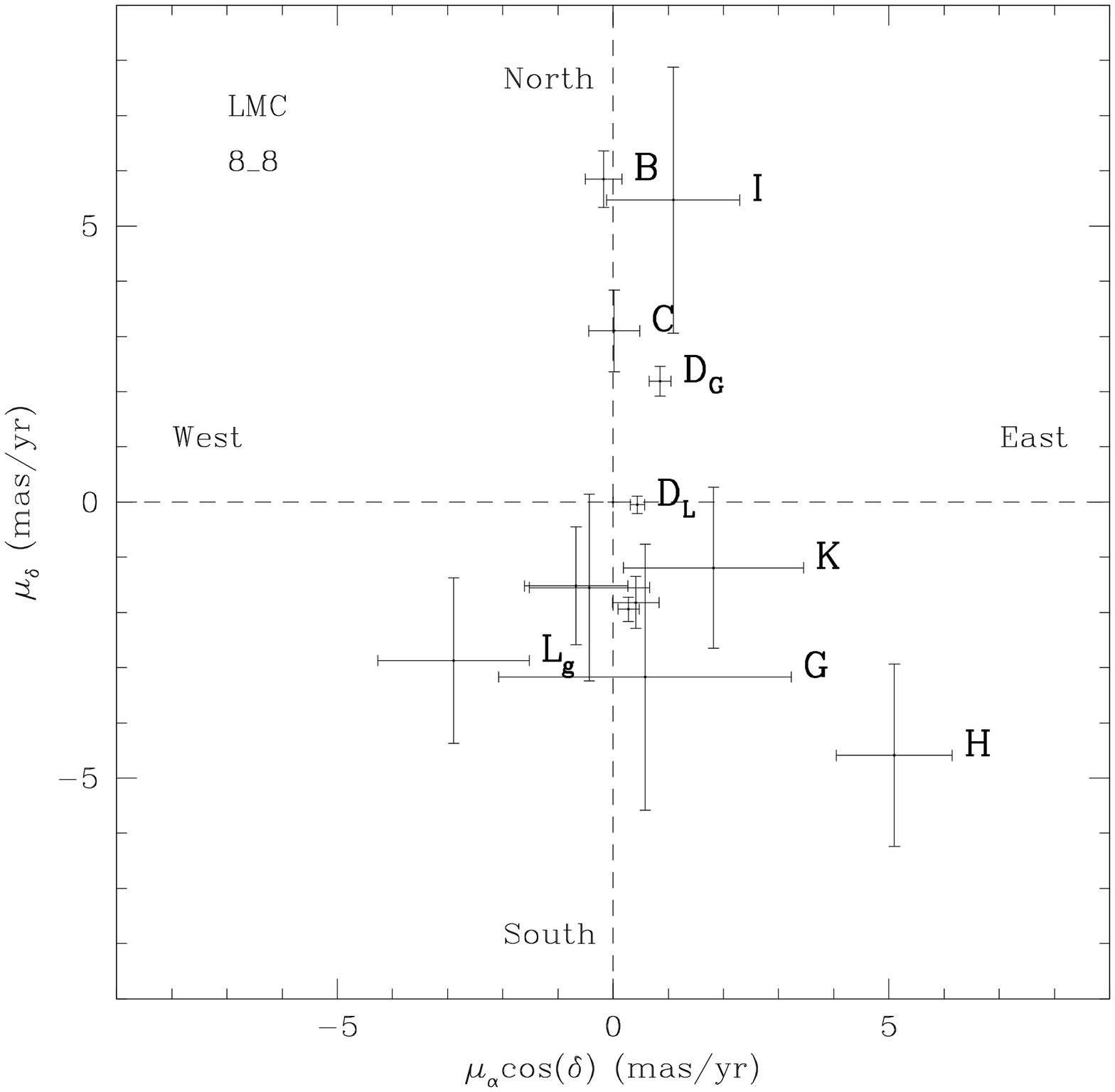}
\includegraphics{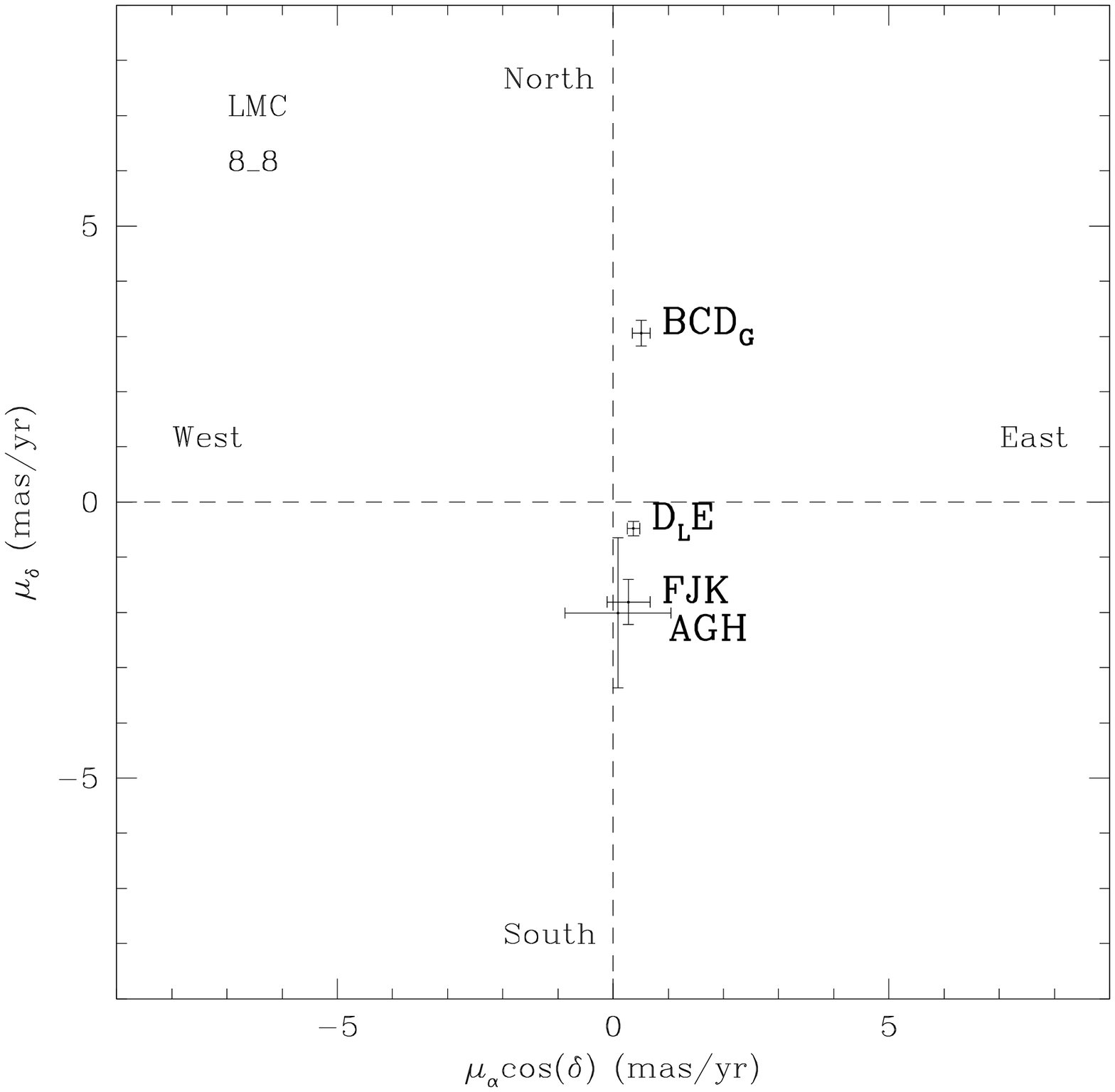}}
\caption{Proper motion derived from the comparison between VMC and 2MASS data in tile LMC $8\_8$. (left) Proper motion values for stars in different CMD regions as identified in Fig. \ref{cmd} and  listed in Table \ref{tab88}. Note that labels for regions A, E, F and J have been omitted for clarity. The point corresponding to region L$_\mathrm{s}$ is outside the range shown in the diagram. (right) Proper motion values for stars in combined regions as listed in Table \ref{tab88avg} dominated respectively by LMC RGB stars (D$_\mathrm{L}$E), AGB stars (FJK), young stars (AGH) and MW foreground stars (BCD$_\mathrm{G}$).}
\label{pm88}
\end{figure*}

\begin{table}
\caption{Relative VMC-2MASS proper motions in tile LMC $8\_8$.}
\scriptsize
\label{tab88}
\[
\begin{array}{lrrrrrrrr}
\hline \hline
\noalign{\smallskip}
\mathrm{CMD}  & \mathrm{N} & \multicolumn{3}{c}{\mu_\alpha cos(\delta)}  & \mathrm{N} & \multicolumn{3}{c}{\mu_\delta} \\
\mathrm{Region} & & \multicolumn{3}{c}{\mathrm{(mas\,yr^{-1})}} & & \multicolumn{3}{c}{\mathrm{(mas\,yr^{-1})}} \\   
 & & \mu & \delta_\mu & \sigma & & \mu & \delta_\mu & \sigma \\   
\hline
\noalign{\smallskip}
\mathrm{A} & 24 & -0.43 & 1.09 & 5.33  & 24 & -1.55 & 1.69 & 8.26 \\
\mathrm{B} & 564 & -0.17 & 0.33 & 7.95 & 556 & +5.85 & 0.51 & 11.92 \\
\mathrm{C} & 338 & +0.02 & 0.46 & 8.49 & 347 & +3.10 & 0.74 & 13.73 \\
\mathrm{D}_\mathrm{L} & 3770 &  +0.44 & 0.13 & 7.88 & 3747 & -0.05 & 0.16 & 9.56 \\ 
\mathrm{D}_\mathrm{G} & 1297 & +0.85 & 0.20 & 7.15 & 1307 & +2.19 & 0.27 & 9.78 \\
\mathrm{E} & 1002 &  +0.28 & 0.19 & 5.93 & 987 & -1.94 & 0.22 & 6.93 \\
\mathrm{F} & 158 & +0.41 & 0.42 & 5.33 & 160 & -1.82 & 0.47 & 5.97 \\
\mathrm{G} & 5 & +0.58 & 2.65 & 5.93 & 5 & -3.17 & 2.41 & 5.40 \\
\mathrm{H} & 2 & +5.10 & 1.05 & 1.48 & 2 & -4.59 & 1.65 & 2.33 \\
\mathrm{I} & 78 & +1.09 & 1.21 & 10.65 & 82 & +5.47 & 2.41 & 21.86 \\
\mathrm{J} & 44 & -0.67 & 0.94 & 6.23 & 44 & -1.52 & 1.07 & 7.11 \\
\mathrm{K} & 12 &  -1.82 & 1.63 & 5.66 & 12 & -1.19 & 1.46 & 5.06 \\
\mathrm{L_s} & 19 & +0.87 & 3.36 & 14.27 & 19 & +12.86 & 4.66 & 20.30 \\
\mathrm{L_g} & 77 & -2.89 & 1.37 & 11.96 & 77 & -2.87 & 1.50 & 13.16 \\
\noalign{\smallskip}
\hline
\end{array}
\]
\end{table}

The PMs of different types of stars are shown in Fig.~\ref{pm88} (left panel). Region B is dominated by MW stars which show a PM clearly distinct from those of the regions dominated by LMC stars. Region I shows a similar motion suggesting that this group may be also dominated by MW dwarfs and that the contribution of LMC supergiants is minor. Stars in regions B, C, D$_\mathrm{G}$ and I have a motion that is systematically different from those in the regions with mostly LMC stars. Low number statistics ($<30$ objects) render PM determinations in regions A, G, H and K highly uncertain. 
The smallest PM uncertainties are for regions D$_\mathrm{L}$ and $E$, that contain a large number of RGB stars, as well as for MW stars in region D$_\mathrm{G}$. The PM of objects morphologically classified as stars in region L, i.e. L$_\mathrm{s}$ is also highly uncertain probably due to a low number statistics. These objects could be compact background galaxies, e.g. quasars, or a minority of RGB stars of the LMC scattered to region L because of their larger extinction. Their PM indicates that they are unlikely LMC stars and since it also differs considerably from the PM of objects morphologically classified as galaxies (L$_\mathrm{g}$), in the same CMD region, the most probable explanation is that they are MW stars (see Sect. \ref{milkyway}). AGB stars populating regions F, J and K show PM values consistent with each other. The PM of MW stars decreases in $\delta$ from region B to C and D$_\mathrm{G}$, unless this effect is due to the $\sim20$\% contamination by LMC stars within the regions. This difference suggests that the D$_\mathrm{G}$ sample is dominated by stars further away.

The right panel of Fig. \ref{pm88} shows the PMs in combined regions of the CMD dominated by similar stars. In region BCD$_\mathrm{G}$ there are mostly MW stars and their PM is directed North and well separated from the motion of other types of stars in the LMC. Region AGH is dominated by LMC young stars (O-type stars and supergiants) with a large uncertainty in its PM value due to the low number of stars. Intermediate-age AGB stars from region FJK show a PM that is distinct from that of older RGB stars (region DL$_\mathrm{E}$) but consistent, within the uncertainties, with that of young stars while RGB stars show the smallest PM value among the LMC types. The values of the PM in the combined regions are presented in Table \ref{tab88avg} (left). The PM of the LMC in this field, with respect to $77$ background galaxies, is $\mu_\alpha cos(\delta)= +3.28\pm0.10$ (stat) $\pm1.37$ (sys) and $\mu_\delta = +2.34\pm0.13$ (stat) $\pm1.50$ (sys) mas yr$^{-1}$. This PM value results from subtracting the L$_\mathrm{g}$ values from the LMC$_\mathrm{8\_8}$ values as listed in Table \ref{tab88} and the formal systematic errors correspond to the uncertainties measured for the galaxies used as calibrators. These errors may be underestimated due to additional effects. Conservative systematic errors are dominated by the 2MASS positional systematics and lead to a maximum systematic PM error of $\sim9$ mas yr$^{-1}$. This value results from the sum in quadrature of the uncertainty of 2MASS measurements ($80$ mas) and a residual systematic pattern in VISTA tiles ($20$ mas), scaled by the time baseline ($10$ years), as well as a residual radial distortion in the current VMC data set ($3-4$ mas yr$^{-1}$). The latter has been estimated from the full radial distortion ($100$ mas) as follows. Let $\Delta$ be the difference between the mean position of all LMC objects (those in CMD regions A, D$_\mathrm{L}$, E, F, G, H, J, K) and background galaxies (region L$_\mathrm{g}$), then the residual radial distortion ($\mu_r$) in units of mas yr$^{-1}$ is given by $\mu_r=(\Delta*100) /(T*r)$ where $r$ is the radius of the field and $T$ is the time that separates the VMC and 2MASS measurements. For $r=1-1.5$ deg and $T=10$ yr we obtain $\mu_r=3-4$ mas yr$^{-1}$. 

\begin{table*}
\caption{Relative combined proper motion in tile LMC $8\_8$ for sources detected in 2MASS All-Sky and $6\times$ 2MASS.}
\label{tab88avg}
\[
\begin{array}{l|rrrrrrrr|rrrrrrrr}
\hline \hline
\noalign{\smallskip}
 & \multicolumn{8}{|c}{\mathrm{VMC - 2MASS}}  & \multicolumn{8}{|c}{\mathrm{VMC - VMC}} \\
\mathrm{CMD}  & \mathrm{N} & \multicolumn{3}{c}{\mu_\alpha cos(\delta)} & \mathrm{N}  & \multicolumn{3}{c|}{\mu_\delta} &  \mathrm{N} & \multicolumn{3}{c}{\mu_\alpha cos(\delta)} & \mathrm{N} & \multicolumn{3}{c}{\mu_\delta} \\
\mathrm{Region}  &  & \multicolumn{3}{c}{(\mathrm{mas}\,\mathrm{yr}^{-1})}  & & \multicolumn{3}{c|}{(\mathrm{mas}\,\mathrm{yr}^{-1})} &  & \multicolumn{3}{c}{(\mathrm{mas}\,\mathrm{yr}^{-1})}  & &  \multicolumn{3}{c}{(\mathrm{mas}\,\mathrm{yr}^{-1})} \\
 & & \mu & \delta_\mu & \sigma & & \mu & \delta_\mu & \sigma & & \mu & \delta_\mu & \sigma & & \mu & \delta_\mu & \sigma \\
\hline
\noalign{\smallskip}
\mathrm{BCD}_\mathrm{G} & 2208 & +0.51 & 0.16 & 7.53 & 2200 & +3.06 & 0.23 & 10.73 & 2167 & +0.88 & 0.15 & 6.88 & 2166 & +4.05 & 0.22 & 10.24 \\
\mathrm{D}_\mathrm{L}\mathrm{E} & 4841 & +0.37 & 0.11 & 7.41 & 4800 & -0.48 & 0.13 & 8.99 & 4772 & +0.45 & 0.08 & 5.25 & 4683 & +0.41 & 0.10 & 7.17 \\
\mathrm{AGH} & 31 & +0.09 & 0.96 & 5.33 & 31 & -2.01 & 1.36 & 7.57 & 31 & -0.54 & 1.17 & 6.54 & 30 & +0.02 & 1.80 & 9.86 \\
\mathrm{FJK} & 216 & +0.28 & 0.39 & 5.72 & 216 & -1.81 & 0.41 & 5.96 & 217 & -0.46 & 0.42 & 6.19 & 212 & -0.80 & 0.48 & 7.05 \\  
\noalign{\smallskip}
\hline
\noalign{\smallskip}
\mathrm{LMC}_{8\_8} & 5006 & +0.39 & 0.10 & 7.33 & 4969 & -0.53 & 0.13 & 8.88 & 4878 & +0.40 & 0.07 & 5.27 & 4786 & +0.34 & 0.10 & 7.15 \\
\noalign{\smallskip}
\hline
\end{array}
\]
\end{table*}

\begin{table*}
\caption{Relative combined proper motion in tile LMC $8\_8$ for sources detected in 2MASS All-Sky.}
\label{tab88avg1}
\[
\begin{array}{l|rrrrrrrr|rrrrrrrr}
\hline \hline
\noalign{\smallskip}
 & \multicolumn{8}{|c}{\mathrm{VMC - 2MASS}}  & \multicolumn{8}{|c}{\mathrm{VMC - VMC}} \\
\mathrm{CMD}  & \mathrm{N} & \multicolumn{3}{c}{\mu_\alpha cos(\delta)} & \mathrm{N} & \multicolumn{3}{c|}{\mu_\delta} &  \mathrm{N} & \multicolumn{3}{c}{\mu_\alpha cos(\delta)} & \mathrm{N} & \multicolumn{3}{c}{\mu_\delta} \\
\mathrm{Region}  &  & \multicolumn{3}{c}{(\mathrm{mas}\,\mathrm{yr}^{-1})}  & & \multicolumn{3}{c|}{(\mathrm{mas}\,\mathrm{yr}^{-1})} &  & \multicolumn{3}{c}{(\mathrm{mas}\,\mathrm{yr}^{-1})} & & \multicolumn{3}{c}{(\mathrm{mas}\,\mathrm{yr}^{-1})} \\
 & & \mu & \delta_\mu & \sigma & & \mu & \delta_\mu & \sigma & & \mu & \delta_\mu & \sigma & & \mu & \delta_\mu & \sigma \\
\hline
\noalign{\smallskip}
\mathrm{BCD}_\mathrm{G} & 1553 & -0.07 & 0.19 & 7.68 & 1552 & +3.89 & 0.30 & 11.69 & 1532 & +0.74 & 0.19 & 7.59 & 1534 & +5.07 & 0.29 & 11.54 \\ 
\mathrm{D}_\mathrm{L}\mathrm{E} & 2413 & -0.05 & 0.13 & 6.35 & 2364 & -1.48 & 0.16 & 7.64 & 2534 & +0.36 & 0.10 & 4.91 & 2311 & -0.41 & 0.14 & 6.51 \\
\mathrm{AGH} & 19 & -0.26 & 1.26 & 5.49 & 19 & -1.66 & 1.74 & 7.60 & 19 & +0.45 & 1.65 & 7.21 & 19 & -1.37 & 3.55 & 15.49 \\
\mathrm{FJK} & 216 & +0.28 & 0.39 & 5.72 & 216 & -1.81 & 0.41 & 5.96 & 217 & -0.46 & 0.42 & 6.19 & 212 & -0.80 & 0.48 & 7.05 \\
\noalign{\smallskip}
\hline
\noalign{\smallskip}
\mathrm{LMC}_{8\_8} & 2580 & +0.02 & 0.12 & 6.30 & 2532 & -1.50 & 0.15 & 7.53 & 2521 & +0.30 & 0.10 & 5.03 & 2474 & -0.45 & 0.13 & 6.58\\
\noalign{\smallskip}
\hline
\end{array}
\]
\end{table*}

\subsection{Inhomogeneous distribution of $6\times$ 2MASS data}
\label{inom}
In case of residual spatial distortions in the VMC and/or 2MASS data the uneven coverage of tile LMC $8\_8$ from the $6\times$ 2MASS catalogue may influence the PM result. To investigate this effect we have re-calculated the PMs excluding the $6\times$ 2MASS data. The number of sources in regions A, D$_\mathrm{G}$, D$_\mathrm{L}$ and L is reduced by about a half and the PM values approach zero in the $\alpha$ direction while the change in the $\delta$ direction enhances the difference between MW and LMC PMs (see Table \ref{tab88avg1}). The latter is probably a result of the reduced contamination between the two classes of stars as selected from the CMD. In this case the PM of young, intermediate-age and old LMC stars are consistent with each other. Because the variations in $\mu_\delta$ occur in opposite directions, for MW and LMC stars, and the fact that at this brightness there are only a handful of background galaxies left, it is not possible to calibrate the PM values to an absolute system that is independent from the inhomogeneous coverage of the field by the $6\times$ 2MASS data. Furthermore, in tile LMC $8\_8$ there are only candidate quasars and they have not yet been spectroscopically confirmed \citep{2013A&A...549A..29C}.

\subsection{Distribution of different types of sources}
\label{uneven}

\begin{figure}
\resizebox{\hsize}{!}{\includegraphics{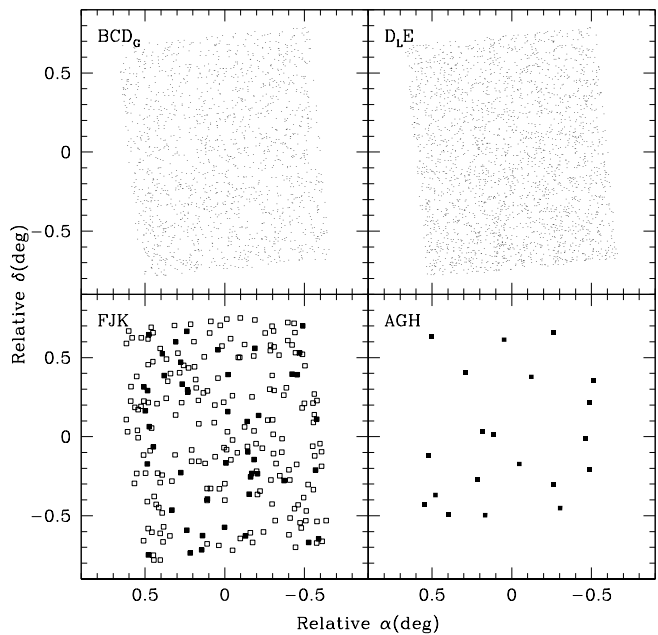}}
\caption{Spatial distribution of different types of sources. (top-left) MW foreground stars. (top-right) LMC RGB stars. (bottom-left) LMC AGB stars where carbon stars are indicated with filled squares. (bottom-right) LMC young stars. The CMD regions to which these objects  belong are indicated in the upper left corner of each panel. The centre of each panel is at $(\alpha_0$, $\delta_0)=(-89^\circ.8464$, $-66^\circ.3413)$.}
\label{sources}
\end{figure}

An uneven coverage of tile LMC $8\_8$ from the different types of sources together with the presence of residual distortions may influence the measurements of the PM. The residual distortion of $\pm100$ mas found in the VMC data do not influence the PM results if the sources are symmetrically distributed around the tile centre. Figure \ref{sources} shows that both RGB stars and MW foreground stars are homogeneously sampling the entire VMC tile and the same is true for AGB stars and young stars despite their reduced number. Hence, we conclude that residual distortions are not responsible for the difference among the PMs of different stellar populations.

\subsection{VMC quality flags and bright stars}
\label{bright}
The quality flags that dominate VMC stars detected in 2MASS correspond to ppErrBits $0-16$ which include stars de-blended by the pipeline. Note that many stars of the Magellanic Clouds have these ppErrBits values because of the high level of crowding of the stellar field. There is only a negligible number of sources affected by bad pixels, low confidence in the default aperture and by a photometric calibration probably subject to photometric errors. Sources located in detector \#16 (with varying quantum efficiency that influences source magnitudes because of difficulties in flat-fielding -- this effect is small at $K_\mathrm{s}$\footnote{http://apm49.ast.cam.ac.uk/\-surveys-projects/\-vista/\-technical/\-known-issues}) and in the underexposed ``wings" of VISTA tiles\footnote{See http://www.vista.ac.uk/} are also included in the analysis. If they were to produce PMs considerably different from those of the other sources in the CMD regions where they are located they would be filtered out by our clipping criteria. Furthermore, because of the brightness of the 2MASS stars at least $7$ epochs of VMC data are always available both in the centre and in the wings of VISTA tiles. For more details about the VISTA quality flags see http://horus.roe.ac.uk/vsa/ppErrBits.html. 

Stars close to the VMC saturation limit have magnitudes $K_\mathrm{s}<11.5$  mag \citep{2011A&A...527A.116C} and ppErrBits $\sim 65000$. A specific pipeline procedure, designed by Irwin \citep{2009UKIRT.25.15}, allows to recover the stellar parameters for sources up to a few magnitude brighter than this limit, hence these stars are included in the analysis of the PM (Fig. \ref{cmd}), but their PM should be taken with care since it may still be influenced by their large brightness. Excluding these bright sources would reduce the number of stars in regions B, C, I and F and remove most of the stars in regions H, G, J and K making the PM in these latter regions highly unreliable. The PMs in the other regions as well as in the combined regions of the CMD, however, are only marginally reduced in $\mu_\delta$ and unaltered in $\mu_\alpha cos(\delta)$, in the VMC-2MASS combination, reflecting the larger influence by LMC stars within the MW regions. The PMs from the VMC data alone stay the same.

\begin{figure}
\resizebox{\hsize}{!}{\includegraphics{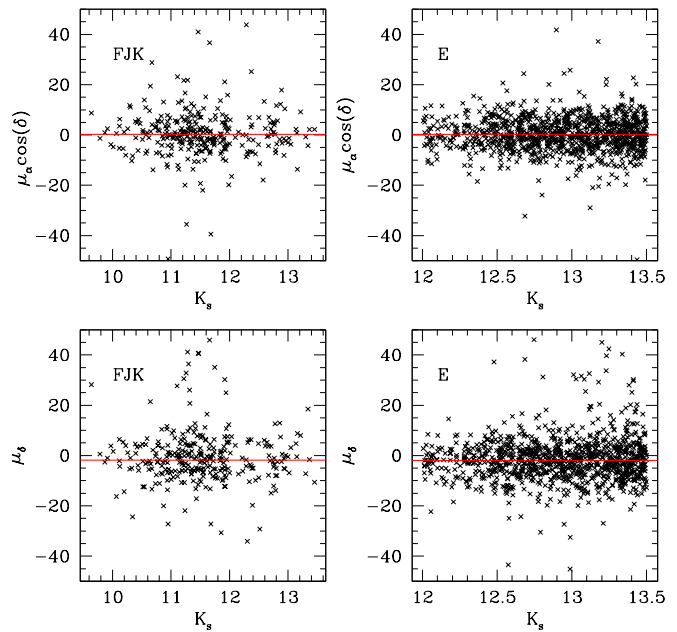}}
\caption{Proper motion as function of $K_\mathrm{s}$ magnitude for AGB stars (regions FJK - left) and RGB stars (region E - right). Horizontal lines indicate the measured PMs (see Tables \ref{tab88} and \ref{tab88avg}).}
\label{new}
\end{figure}

In Fig. \ref{new} the PM of AGB stars, from regions FJK in Table \ref{tab88avg}, is shown next to the PM of RGB stars, from region E only in Table \ref{tab88}. All distributions have a similar dispersion indicating that the inclusion of bright stars does not significantly alter the measured PMs.

\section{VMC-VMC proper motion}
\label{vpm}

\subsection{2MASS stars}
\label{vmcvmc}
In order to evaluate if the comparison between VMC and 2MASS data is affected by systematics due to the different instruments we have examined the PMs that can be derived from VMC data alone for the same objects cross-correlated with the 2MASS data. For tile LMC $8\_8$ VMC data provide a time baseline of $377$ d and $11$ individual deep epochs. We found, however, that the observations obtained in November 2009 are affected by a systematic error of $\sim10$ mas, these epochs are therefore excluded which leave a VMC time baseline of $336$ d and $7$ deep epochs. The reason for this systematic shift, which affects only individual tile coordinates and not deep tile coordinates, is unknown but a re-calculation of the astrometric solution in the making of the tiles removes it. The corrected data will be included in subsequent VMC releases. The PM has been derived from the slope of the linear fit to the individual displacements. This method produces the best fit line because the expected relation between displacement and time is linear, the independent variable (time) is measured without errors, the errors on the displacements are symmetric, have a similar dispersion and are independent from epoch to epoch \citep[e.g.][]{1990ApJ...364..104I}. Results for the combined CMD regions, after the rejection of outliers as in Sect. \ref{pm}, are listed in Tables \ref{tab88avg} and \ref{tab88avg1} (right).

\begin{figure}
\resizebox{\hsize}{!}{\includegraphics{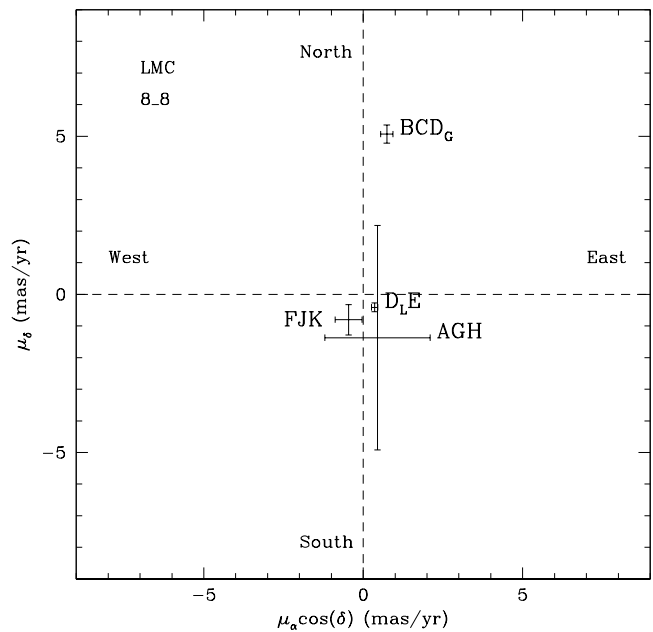}}
\caption{Same as Fig.~\ref{pm88} (right) but for proper motions derived using only VMC data for sources in the 2MASS All-Sky sample.}
\label{vmcpm}
\end{figure}

Despite the reduced time baseline, the uncertainties in the VMC-VMC PMs are comparable with those from the VMC-2MASS PMs. Figure \ref{vmcpm} shows the PM from VMC-VMC data for sources in the 2MASS All-Sky sample. The PM for MW stars (region BCD$_\mathrm{G}$) is larger than in Fig. \ref{pm88} probably because of the reduced contamination by LMC stars (Sect. \ref{inom}), i.e. there is less overlap between the MW and LMC branches at $K_\mathrm{s}<14.3$ mag. Except for the PM of AGB stars (region FJK) Figs. \ref{vmcpm} and \ref{pm88} show consistent results for LMC stars. AGB stars are among the brightest sources detected by the VMC survey and it is possible that this influences the determination of their centroid position. A comparison between the PM of VMC-2MASS and VMC-VMC for sources in the 2MASS All-Sky sample only shows a systematic difference for both MW and LMC stars where, however, the relative distance is conserved and amounts to $\mu_\delta \sim 5.3$ mas yr$^{-1}$ (Table \ref{tab88avg1}). Contrary to the situation where sources from the $6\times$ 2MASS sample are also included, there are not enough background galaxies, or quasars, to anchor the PMs to an absolute system (see Sect. \ref{inom}). 

We conclude that the homogeneity of the VMC data across the tile provides reliable PMs. The selection of a sample of LMC stars as clean as possible from MW stars compensates the influence of the short VMC time baseline with respect to a sample originally contaminated but for which the PM can be evaluated across a time baseline about $10$ times larger, cf. LMC$_\mathrm{8\_8}$ PM from Table \ref{tab88avg} (left) and Table \ref{tab88avg1} (right).

\subsection{VMC stars}
\label{vmcvmcsection}
\begin{figure}
\resizebox{\hsize}{!}{\includegraphics{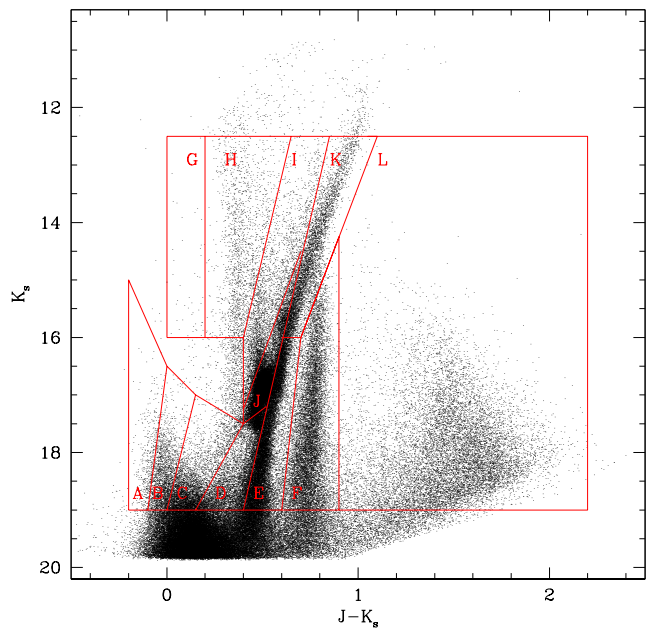}}
\caption{CMD of VMC sources in tile LMC $8\_8$. Region boundaries are defined as in Table \ref{vpop} to distinguish among stars with a different mean age.}
\label{vmcbox}
\end{figure}

A considerable statistical leap forward in measuring PMs is obtained using VMC sources fainter than the 2MASS limit. These include sources that span the whole extent of the RGB, red clump giant stars, main sequence stars and a large number of both MW and background galaxies.  Figure \ref{vmcbox} shows the distribution of $151$,$724$ VMC sources that are detected in three wave bands, with photometric errors $<0.1$ mag and ppErrBits $0-16$ in each band. Note that sources located in detector \#16 and in the wings of VISTA tiles are excluded. Contrary to the 2MASS stars, it is more likely that faint VMC stars in the wings would not have all $7$ $K_\mathrm{s}$ epochs producing an uneven sampling of the bright and faint VMC stars across the tile. At faint magnitudes there will also be many more sources in detector \#$16$ that may influence the PM of a given CMD region and these are also excluded from the subsequent analysis.
We have added $0.001$ to $K_\mathrm{s}$ and $0.021$ to $J$ to express VISTA magnitudes into the Vega system. The region boundaries and the mean age of the stars within each region are listed in Table \ref{vpop}. They have been established according to the analysis of the star formation history in tile LMC $8\_8$ by \citet{2012A&A...537A.106R}. The faint limit includes all sources with photometric uncertainties in the $K_\mathrm{s}$ band $<0.1$ mag at all epochs (see Sect. \ref{vmcdata}) while the bright limit aims at including a large statistical sample well below the level of bright stars (see Sect. \ref{bright}) and with a moderate overlap with the faint 2MASS magnitudes; the tip of the RGB is at $K_\mathrm{s}\sim 12$ mag.

\begin{figure}
\resizebox{\hsize}{!}{\includegraphics{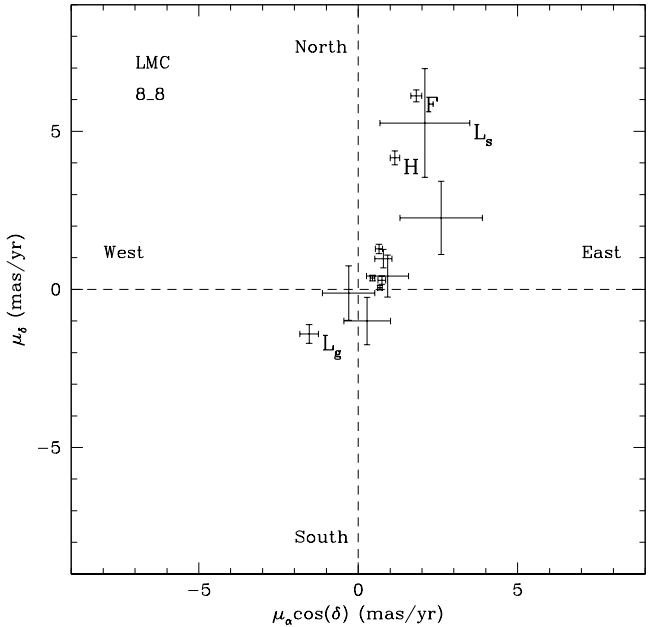}}
\caption{Proper motion derived from VMC data in tile LMC $8\_8$ for stars in different CMD regions as identified in Fig. \ref{vmcbox} and listed in Table \ref{vtab88}. Note the clear separation between MW stars (F, H and L$_\mathrm{s}$) and LMC stars as well as background galaxies (L$_\mathrm{g}$). Note also that points corresponding to LMC stars are not labelled for clarity.}
\label{faintpm}
\end{figure}

Following the same procedure as described in Sect. \ref{vmcvmc} we have derived the mean PM of stars in each CMD region and the results are shown in Fig. \ref{faintpm} and listed in Table \ref{vtab88}. The PM of MW stars (regions F and H) is clearly distinct from those of LMC stars and background galaxies (region L$_\mathrm{g}$). Furthermore by distinguishing between objects with stellar-like morphologies and objects with galaxy-like morphologies it is possible to separate the PM of late-type MW dwarf stars in region L obtaining an L$_\mathrm{s}$ value consistent with PM values of stars in regions F and H dominated by MW stars. The PM of LMC stars (regions B, C, D, E, G, I, J, K) are clustered around zero but for region A  which appears to deviate from this despite the large uncertainties. Among the best determined LMC PM values, the PM of region I differs from those of regions E, J and K. The PM of region D is consistent with that of region I within the uncertainties. The PM of region C is consistent with both the PM of regions D, I and E, J, K within the uncertainties. The PM of the LMC stars as a whole, calibrated with respect to background galaxies, corresponds to $\mu_\alpha cos(\delta) = +2.20\pm0.06$ (stat) $\pm0.29$ (sys) and $\mu_\delta = +1.70\pm0.06$ (stat) $\pm0.30$ (sys) mas yr$^{-1}$. This PM value results from subtracting the L$_\mathrm{g}$ values from the LMC$_\mathrm{8\_8}$ values as listed in Table \ref{vtab88} and the formal systematic errors correspond to the uncertainties measured for the galaxies used as calibrators. Similarly to Sect. \ref{pm} these errors may be underestimated due to a residual systematic pattern in the VISTA data ($20$ mas) and a residual radial distortion in the current VMC data set ($4-5$ mas). The latter has been estimated as in Sect. \ref{pm} for $T=1$ yr. Compared to the PM values derived in Sect. \ref{pm} there is good agreement within the formal uncertainties which are reduced here thanks to the large sample offered by the VMC data.

\begin{figure}
\resizebox{\hsize}{!}{\includegraphics{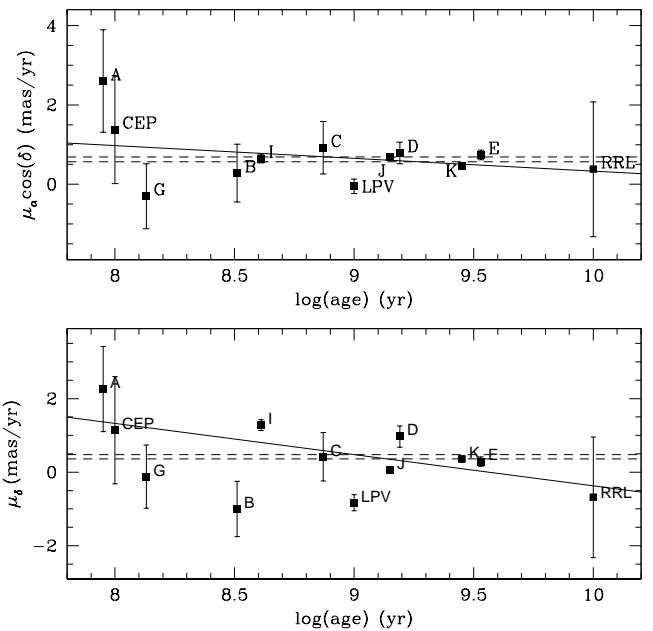}}
\caption{Proper motion as a function of age. Points correspond to the median of the age distribution of stars in each CMD region (Table \ref{vpop}) while ages of variable stars have been assigned as follows: $0.1$ Gyr to Cepheids (CEP), $1$ Gyr to Long Period Variables (LPVs) and $10$ Gyr  to RR Lyrae stars (RRL). Dashed lines represent the uncertainty on the average proper motion from all LMC stars within the tile. Continuous lines indicate the best weighted least square fit line. }
\label{age}
\end{figure}

Figure \ref{age} shows the PM of stars in individual CMD regions as a function of log(age). To determine the ages of stars in each box, we take the best fitting model derived from \citet{2012A&A...537A.106R} for this region. Their SFH is used to produce a simulated CMD including photometric errors and the MW foreground, from which we determine the median ages, their $1\sigma$ confidence interval and the contamination from MW stars. These data are presented in Table \ref{vpop}. In general, there is no significant trend with age but the scatter around the mean PM value is larger in $\mu_\delta$ than in $\mu_\alpha (cos\delta)$. A weighted least square fit, including the PM of variable stars (Sect. \ref{variables}), with weights  corresponding to the reciprocal of the uncertainties squared, gives: 
\begin{equation}
\label{one}
\mu_\alpha cos(\delta) = -0.32\times \mathrm{log(age)} + 3.55
\end{equation}
\begin{equation}
\label{two}
\mu_\delta = -0.85\times \mathrm{log(age)} + 8.11
\end{equation} 
where the rms is $0.17$ and $0.36$ mas yr$^{-1}$ in Eq. \ref{one} and Eq. \ref{two}, respectively. 
At the distance of the LMC $0.2$ mas yr$^{-1}$ corresponds to $\sim50$ km s$^{-1}$ and it is expected that stars populating the outer structure of the LMC would differ by this amount from those located in the disk of the galaxy. 
The PM difference between young ($0.1$ Gyr) and old ($10$ Gyr) stars is significant with respect to the rms. It amounts to $0.64$ mas yr$^{-1}$ in the $\mu_\alpha cos(\delta)$ direction and to $1.70$ mas yr$^{-1}$ in the $\mu_\delta$ direction which correspond to a difference of $\sim1.8$ mas yr$^{-1}$ ($\sim450$ km s$^{-1}$) in a North-West direction consistent with the clockwise rotation of the galaxy.
This value is most likely too high since line-of-sight velocities of red supergiant stars in the region are $\sim300$ km s$^{-1}$ \citep{2003AJ....126.2867M}, but suggests that we have measured the PM of stars that are clearly associated to different sub-structures within the LMC.  In addition, studies of the line-of-sight kinematics indicate that the young stars rotate faster than the old stars, and have a smaller velocity dispersion \citep[e.g.][]{2013arXiv1305.4641V}; a component of this motion is also seen in the PM direction. 
A thorough investigation of PM variations as a function of age and of the internal kinematics of the LMC will be explored in more detail when more VMC tiles will be analysed. 

\begin{table}
\caption{Relative VMC-VMC proper motions in tile LMC $8\_8$.}
\scriptsize
\label{vtab88}
\[
\begin{array}{lrrrrrrrr}
\hline \hline
\noalign{\smallskip}
\mathrm{CMD}  & \mathrm{N} & \multicolumn{3}{c}{\mu_\alpha cos(\delta)}  & \mathrm{N} & \multicolumn{3}{c}{\mu_\delta} \\
\mathrm{Region} & & \multicolumn{3}{c}{\mathrm{(mas\,yr^{-1})}} & & \multicolumn{3}{c}{\mathrm{(mas\,yr^{-1})}} \\   
 & & \mu & \delta_\mu & \sigma & & \mu & \delta_\mu & \sigma \\   
\hline
\noalign{\smallskip}
\mathrm{A} & 250 & +2.60 & 1.29 & 20.43 & 249 & +2.26 & 1.16 & 18.30 \\
\mathrm{B} & 790 & +0.28 & 0.73 & 20.43 & 790 & -1.00 & 0.75 & 21.00 \\
\mathrm{C} & 950 & +0.92 & 0.66 & 20.22 & 949 & +0.42 & 0.66 & 20.43 \\
\mathrm{D} & 3428 & +0.79 & 0.27 & 16.06 & 3444 & +0.97 & 0.29 & 16.92 \\
\mathrm{E} & 11182 & +0.74 & 0.12 & 12.96 & 11256 & +0.29 & 0.13 & 13.78 \\
\mathrm{F} & 7104 & +1.82 & 0.17 & 14.06 & 7156 & +6.12 & 0.19 & 16.18 \\
\mathrm{G} & 47 & -0.30 & 0.82 & 5.62 & 45 & -0.12 & 0.86 & 5.79 \\
\mathrm{H} & 1583 & +1.15 & 0.15 & 5.96 & 1568 & +4.16 & 0.22 & 8.52 \\
\mathrm{I} & 3239 & +0.65 & 0.11 & 6.47 & 3227 & +1.28 & 0.15 & 8.40 \\
\mathrm{J} & 17042 & +0.68 & 0.07 & 8.76 & 17069 & +0.05 & 0.07 & 9.55 \\
\mathrm{K} & 5436 & +0.45 & 0.07 & 5.00 & 5352 & +0.36 & 0.09 & 6.75 \\
\mathrm{L_{s}} & 229 & +2.09 & 1.41 & 21.41 & 231 & +5.26 & 1.72 & 26.08 \\
\mathrm{L_{g}} & 7834 & -1.54 & 0.29 & 26.05 & 7865 & -1.41 & 0.30 & 26.84 \\
\hline
\noalign{\smallskip}
\mathrm{LMC}_{8\_8} & 36468 & +0.66 & 0.06 & 10.71 & 36662 & +0.29 & 0.06 & 11.65 \\
\noalign{\smallskip}
\hline
\end{array}
\]
\end{table}

\subsubsection{Colour, magnitude and position effects}
The different atmospheric refraction that stars of different colour experience may induce systematic shifts in the PM measurements. Although we are restricting our analysis to the $K_\mathrm{s}$ band, we are comparing the PM of stars with different colours. To investigate this effect we have analysed the PM as a function of colour within each region as defined in Fig. \ref{vmcbox} and found no significant trend. Similarly, there is no trend with magnitude. Figure \ref{tmagcollmc} shows as an example the corresponding diagrams for LMC stars included in region $ABCDEGHIJK$ of Fig. \ref{vmcbox}, which sample a range of $6$ in magnitude and $1$ in colour.

Furthermore, we have analysed the PM as function of position, right ascension and declination, within the tile and found that despite oscillations, especially of $\mu_\delta$ as a function of $\delta$, there is no significant trend within the uncertainty associated to the PM mean values. This effect, that is probably due to residual astrometric distorsions among the VISTA detectors, affects equally all stars homogeneously distributed within the tile and appears symmetric with respect to the centre of the distributions, hence it does not influence the comparison among their PMs. Nevertheless, we plan to investigate in more detail the origin of this effect and eventually develop a way to correct for it following the analysis of more VISTA tiles.

\subsubsection{Variable stars}
\label{variables}
\begin{figure}
\resizebox{\hsize}{!}{\includegraphics{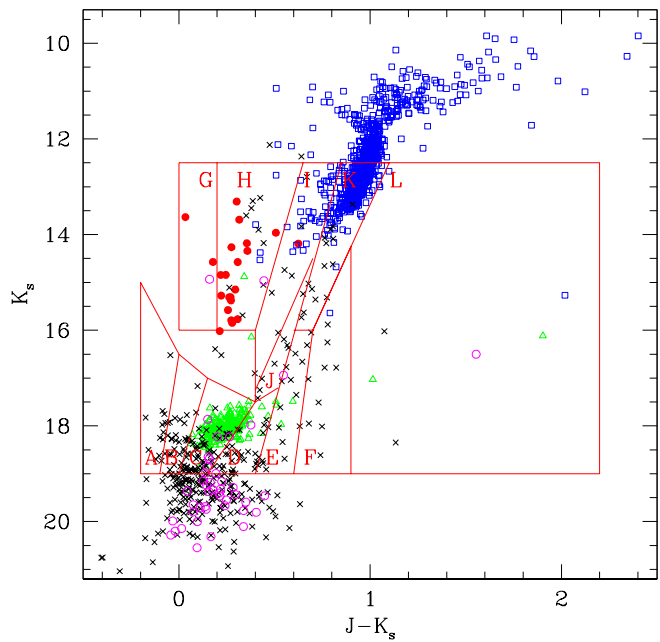}}
\caption{CMD of variables stars in tile LMC $8\_8$. The different symbols indicate LPVs (empty squares -- blue), Classical Cepheids (filled circles -- red), RR Lyrae stars (empty triangles -- green), Eclipsing Binaries (crosses -- black) and $\delta$ Scuti stars (empty circles -- magenta), respectively. Region boundaries are as in Fig. \ref{vmcbox}.}
\label{cmdvar}
\end{figure}

Tile LMC $8\_8$ contains $\sim1500$ known variable stars: $22$ Classical Cepheids, $44\, \delta$ Scuti stars, $364$ Eclipsing Binaries, $844$ Long Period Variables (LPVs) and $221$ RR Lyrae stars (RRL) where the classification results from the analysis of their OGLE or EROS2 light-curve \citep{2012AcA....62..219S, 2013arXiv1310.6849M}. Figure \ref{cmdvar} shows the distribution of variable stars in the CMD with superimposed the regions defined in Sect. \ref{vmcvmcsection} to distinguish among stars of a different age. Most RR Lyrae stars are included in region C and Cepheids in region H while O-rich LPVs occupy mostly region K but many are also present in region I. Eclipsing Binary stars are distributed in several regions and many of them have $K_\mathrm{s}<19$ mag like the majority of $\delta$ Scuti stars, hence are outside the area used to derive the PM; their photometric uncertainty at individual $K_\mathrm{s}$ epochs is $>0.1$ mag.

\begin{figure}
\resizebox{\hsize}{!}{\includegraphics{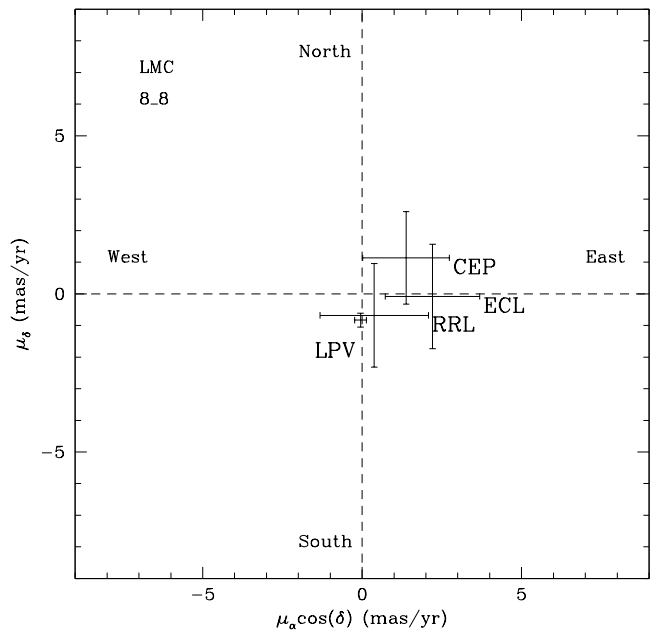}}
\caption{Proper motion derived from VMC data in tile LMC $8\_8$ for different types of variable stars as shown in Fig. \ref{cmdvar}.}
\label{pmvar}
\end{figure}

The PM of different types of variable stars is shown in Fig. \ref{pmvar} and listed in Table \ref{vartab}. The PM of Cepheids, Eclipsing Binaries and RR Lyrae stars are consistent with each other within the uncertainties. The PM of LPVs is consistent with that of RR Lyrae stars but may differ from that of Cepheids and Eclipsing Binaries. We consider Cepheids as representative of the LMC population in region H which contains also a large fraction of MW stars. Cepheids are on average $0.1$ Gyr old and their PM agrees with that of young stars in regions A and G (Fig. \ref{age}). RR Lyrae stars sample the oldest stellar population and their age is approximately $10$ Gyr. Their PM is not significantly different from that of younger stars due to the large uncertainties associated to both groups of stars (Fig. \ref{age}). On the other hand, the PM of LPVs, that are on average $1$ Gyr old, differs from the PM of all stars in region K (Fig. \ref{age}). This is probably due to the presence of a numerous population of RGB stars within the region (Fig. \ref{vmcbox}). The class of Eclipsing Binaries spans a large range of ages. Their PM, due to their low number, is in line with that of the other stars (cf. Tables 4 and 5).

\begin{table}
\caption{Relative proper motions for variable stars in tile LMC $8\_8$.}
\footnotesize
\label{vartab}
\[
\begin{array}{l|rrrr|rrrr}
\hline \hline
\noalign{\smallskip}
\mathrm{CMD}^\mathrm{a}  & \mathrm{N} & \multicolumn{3}{c|}{\mu_\alpha cos(\delta)}  & \mathrm{N} & \multicolumn{3}{c}{\mu_\delta} \\
\mathrm{Region} & & \multicolumn{3}{c|}{\mathrm{(mas\,yr^{-1})}} & & \multicolumn{3}{c}{\mathrm{(mas\,yr^{-1})}} \\   
 & & \mu & \delta_\mu & \sigma & & \mu & \delta_\mu & \sigma \\   
\hline
\noalign{\smallskip}
\mathrm{CEP} & 22 & +1.38 & 1.36 & 6.39 & 22 & 1.14 & 1.46 & 6.84 \\
\mathrm{LPV} & 815 & -0.05 & 0.18 & 5.06 & 810 & -0.83 & 0.22 & 6.15 \\
\mathrm{RRL} & 170 & +0.38 & 1.70 & 22.15 & 168 & -0.68 & 1.64 & 21.21\\
\mathrm{ECL} & 97 & +2.21 & 1.48 & 14.53 & 96 & -0.08 & 1.65 & 16.12 \\
\hline
\end{array}
\]
$^\mathrm{a}$ Stellar type abbreviations are as in Fig. \ref{age} and ECL stands for Eclipsing Binaries.
\end{table}

\section{Comparison to other PM results}
\label{discussion}

\subsection{LMC proper motion}
There is one measurement of the field PM close to the location of tile LMC $8\_8$ made by \citet{2000AJ....120..845A} and subsequently revised by \citet{2006AJ....131.1461P} and \citet{2011RMxAA..47..333P}. These authors measured a PM of $\mu_\alpha cos(\delta) = 2.1\pm0.2$ and $\mu_\delta = 1.5\pm0.2$ mas yr$^{-1}$ from $11$ observing epochs distributed across a time baseline of $12$ years. Their field centre is at $(\alpha, \delta) = (06$:$15$:$09.1$, $-66$:$17$:$40)$ and their field of view covered approximately $2^\prime \times 3^\prime$. Their PM is the average of the contribution from $16$ different stars: bright main sequence stars, RGB and red clump stars, AGB stars, a few  Cepheids and RR Lyrae stars.

Tile LMC $8\_8$ is located to the West of this region and is displaced by $\sim16^\mathrm{m}$ in $\alpha$. The PM that we derive, with respect to background galaxies, corresponds to $\mu_\alpha cos(\delta) = +2.20\pm0.06\, \mathrm{(stat)}\pm0.29\, \mathrm{(sys)}$ and $\mu_\delta = +1.70\pm0.06\, \mathrm{(stat)}\pm0.30\, \mathrm{(sys)}$ mas yr$^{-1}$ where systematic uncertainties may be reduced if only compact background galaxies are selected \citep[e.g.][]{2013ApJ...768..139S}. This is in excellent agreement with the \citet{2011RMxAA..47..333P} results and suggests that the multi-epoch VMC data are a powerful means to measure the PM of stars in the Magellanic Clouds. The high astrometric accuracy and large statistical sample they offer compensates for the short time baseline available, but the latter may be longer for other tiles. 

These ground-based results disagree with results obtained with the HST \citep{2006ApJ...638..772K, 2008AJ....135.1024P}. A discrepancy of about $2\sigma$ (especially in $\mu_\delta$) was attributed by \citet{2011RMxAA..47..333P}, who observed a common field using the same background quasar as astrometric reference, to residual systematic effects. It is curious though that with a completely independent method we have reproduced the \citet{2011RMxAA..47..333P} results, this means that both studies are either influenced by similar systematics or are free from them. In our study the good agreement of the 2MASS-VMC (Sect. \ref{pm}) and VMC-VMC (Sect. \ref{vmcvmcsection}) PMs within their formal systematic errors may suggest that the real systematics PM errors are not much larger than a few mas yr$^{-1}$ making our result in line with HST measurements. With only one field, however, our pilot investigation is not yet able to address the large implied internal dynamics/overall dynamics, but this will be further addressed once more fields, observed as part of the VMC survey, become available.

\subsection{MW proper motion}
\label{milkyway}

In the direction of tile LMC $8\_8$ there are not PM observations of MW foreground stars that we can compare with our results. Therefore, we have created mock observations using the stellar population synthesis code frame-work GALAXIA \citep{2011ApJ...730....3S}. We use the standard model implemented by the authors, which is based on the popular and well tested Besan\c{c}on model \citep{2003A&A...409..523R}. To derive PMs for all stars we inverted the method presented by \citet{1987AJ.....93..864J} and assumed the local standard of rest to be $226.84$ km s$^{-1}$ (as set by the model) and the solar peculiar motion, $(U, V, W_\odot)$, to be $(11.1, 12.24, 7.25)$ km s$^{-1}$ \citep{2010MNRAS.403.1829S}. Table \ref{mwpm} shows the PM values obtained for MW stars detected by 2MASS occupying regions B, C and D$_\mathrm{G}$ as defined in Fig. \ref{cmd} and for MW stars detected by VMC in regions F and H as defined in Fig. \ref{vmcbox}. The stars in these mock samples have distances predominantly $(90\%)$ below $2$ kpc for the 2MASS stars and below $3$ kpc for the VMC stars.

\begin{table}
\caption{Proper motions of MW stars from Besan\c{c}on models.}
\footnotesize
\label{mwpm}
\[
\begin{array}{cc|rcccc}
\hline \hline
\noalign{\smallskip}
\mathrm{CMD} & \mathrm{Region} & \mathrm{N} & \multicolumn{2}{c}{\mu_\alpha\,cos(\delta) }& \multicolumn{2}{c}{\mu_\delta} \\
 & & & \multicolumn{2}{c}{\mathrm{(mas\,yr^{-1})}} & \multicolumn{2}{c}{\mathrm{(mas\,yr^{-1})}} \\ 
 & & & \mu & \sigma & \mu & \sigma \\
 \hline
\noalign{\smallskip}
\mathrm {Fig.\,\ref{cmd}} & \mathrm{B} & 288 & +1.77 & 0.55 & +10.14 & 1.14 \\
\mathrm {Fig.\,\ref{cmd}} & \mathrm{C} & 159 & +1.41 & 1.21 & +9.98 & 1.70 \\
\mathrm {Fig.\,\ref{cmd}} & \mathrm{D}_\mathrm{G} & 286 & +1.46 & 0.35 & +7.23 & 0.53 \\
\hline
\noalign{\smallskip}
\mathrm {Fig.\,\ref{vmcbox}} & \mathrm{F} & 6488 & +1.16 & 0.07 & +7.32 & 0.31 \\
\mathrm {Fig.\,\ref{vmcbox}} & \mathrm{H} & 809 & +1.64 & 0.18 & +7.04 & 0.13 \\
\hline
\end{array}
\]
\end{table}

With respect to Fig. \ref{pm88} (left) we confirm that MW stars in regions B, C and D$_\mathrm{G}$ show a decreasing PM in $\mu_\delta$. The $D_\mathrm{G}$ sample is on average $0.5$ kpc closer than the $B-C$ sample. 
The $\mu_\delta$ derived from the models are significantly larger than the values derived from the VMC-2MASS PM, even after calibration with respect to background galaxies (Table \ref{tab88}). These differences and also the numbers of MW sources within each region may be due to the contamination by LMC sources in our selection. These effects may also influence differences in $\mu_\alpha cos(\delta)$ but it is also possible that the models are not entirely representative of all of the MW population detected by VMC towards this tile. According to the model the 2MASS MW sample consists mostly of thin disk stars.

For stars in a CMD region almost completely dominated by MW stars (region F of Fig. \ref{vmcbox}) the PM derived from the VMC data with respect to background galaxies, $\mu_\alpha cos(\delta) = +3.36\pm0.17\, \mathrm{(stat)}\pm0.29\, \mathrm{(sys)}$ and $\mu_\delta = +7.53\pm0.19\, \mathrm{(stat)}\pm0.30\, \mathrm{(sys)}$ mas yr$^{-1}$, is in very good agreement with the value obtained from the GALAXIA/Besan\c{c}on models in $\mu_\delta$ but differs in $\mu_\alpha cos(\delta)$ by $\sim2.2$ mas yr$^{-1}$.  The difference is reduced to $\sim1$ mas yr$^{-1}$ for stars in region H but in $\mu_\delta$ the VMC data show a smaller value than that derived from the models. These differences are due to the $\sim25\%$ LMC stars present in region H that reduce the average PM of stars within the region. Since the model underpredicts the mean $\mu_\alpha cos(\delta)$ this discrepancy cannot be explained by a contamination by LMC stars. Using the mean distance of the stars in region F, $1.7$ kpc, we can transform the PMs into a Galactocentric Cartesian coordinate system. This reveals that the model underpredicts the vertical motions, W, of the stars by $15$ km s$^{-1}$ and overpredicts the velocities in the direction of the Galactic rotation, V, by $9$ km s$^{-1}$. The difference in U is less than $1$ km s$^{-1}$. While the difference in V is not unexpected given the large uncertainties in the local standard of rest \citep[e.g.][]{2008ApJ...689.1044G, 2012MNRAS.427..274S}, the larger deviation in W is comparable to the local velocity dispersion in the $z$-direction of the model thin disk \citep[$21$ km s$^{-1}$;][]{2011ApJ...730....3S}. 
In the range of magnitudes $16<K_\mathrm{s}<19$ mag the MW foreground is a mixture of thin and thick disk stars according to the model.

\section{Conclusions}
\label{conclusion}

In this study we derived proper motions for different types of LMC stars from the combination between VMC and 2MASS observations, spanning a time baseline of $\sim10$ years, and using VMC data alone, across a time baseline of about $1$ year. Stars that differ by age were selected from the colour-magnitude diagram, $(J-K_\mathrm{s})$ vs. $K_\mathrm{s}$, and include: most of the RGB stars, red clump giant stars and main sequence stars as well as variable stars like Classical Cepheids, RR Lyrae stars, LPVs and $\delta$ Scuti stars. The proper motion of MW foreground disk stars, including late-type dwarfs, is easily distinguished from that of LMC stars and is in general agreement with the expectations from the Besan\c{c}on MW stellar population models. Background galaxies, detected in large numbers in the VMC data, are used to express the proper motion results into an absolute reference system. We derive that the LMC stars in tile LMC $8\_8$, including the South Ecliptic Pole, have a proper motion corresponding to $\mu_\alpha cos(\delta) = +3.28\pm0.10$ (stat) $\pm1.37$ (sys) and $\mu_\delta = +2.34\pm0.13$ (stat) $\pm1.50$ (sys) mas $^{-1}$ from the 2MASS-VMC combination and of $\mu_\alpha cos(\delta) = +2.20\pm0.06$ (stat) $\pm0.29$ (sys) and $\mu_\delta = +1.70\pm0.06$ (stat) $\pm0.30$ (sys) mas yr$^{-1}$ from the VMC-VMC combination. The latter is in excellent agreement with the previous measurement by \citet{2011RMxAA..47..333P} in a nearby field, $\mu_\alpha cos(\delta) = +2.2\pm0.2$ and $\mu_\delta = 1.5\pm0.2$ mas yr$^{-1}$, but our statistical uncertainties are a factor of three smaller and are directly comparable to uncertainties derived from HST observations. Formal as well as the maximum expected systematic errors, which dominate at present the proper motion error budget, are preliminary estimates and they will be further investigated in subsequent analysis of the data, i.e. they are likely to decrease due to the improved reduction of the VISTA data and the increase of the time baseline. We measure a decrease of the proper motion with age where young stars stretch to the North-East and old stars to the South-West. This difference is linked to both kinematic differences between young and old stars and to different hosting structures. The recent analysis of the rotation of the LMC, combining proper  motion and line-of-sight velocities, confirms that the distribution of stars cannot be explained by a single flat disk in circular rotation and that variations in both stellar population and radius exist \citep{2013arXiv1305.4641V}.

The complete coverage offered by the VMC data across the Magellanic system, and the possibility of distinguishing populations of different types by their offsets in colours and magnitude, produces PMs that are better suited to complement the large spectroscopic surveys such as \citet{2011ApJ...737...29O} who kinematically detected SMC stars captured by the LMC. With the completion of the VMC survey we will be able to check if their differential proper motion is high enough to be detected as well as providing indications for further sub-structures within the Magellanic system. Furthermore, the homogeneous proper motion sampling of the Magellanic system's area, provided by the VMC data, will allow us to analyse in detail the rotation curve and infer the mass of the galaxies. 

\subsection{Future work}
The VMC survey, which was not initially designed to measure proper motions, provides a valuable independent measurement of the proper motion of different types of stars in the Magellanic Clouds. The analysis of just one out of $>100$ tiles available across the Magellanic system foresees the significant impact that these proper motion measurements will have for the study of the internal kinematics of the system and of their absolute proper motion which is a fundamental ingredient in understanding the orbital history of the galaxies. The VMC data provide also an independent measure of the geometry of the galaxies, for different stellar populations (young Cepheids, intermediate-age red clump stars and old RR Lyrae stars),  which is necessary to calculate centre-of-mass proper motions and investigate sub-structures in space and time.

Two subsequent papers will address the LMC and SMC PMs, respectively, using the VMC data already available for a number of tiles across each galaxy.

Uncertainties in the proper motion derived from VMC-VMC data will improve due to the increase of the time baseline, since the survey is expected to last for several more years, and due to a better characterisation of the astrometric reference system with e.g. the use of spectroscopically confirmed quasars \citep{2013ApJ...775...92K}.

\begin{acknowledgements}
MC acknowledges support from the Alexander von Humboldt Foundation. This publication makes use of data products from the Two Micron All Sky Survey, which is a joint project of the University of Massachusetts and the Infrared Processing and Analysis Center/California Institute of Technology, funded by the National Aeronautics and Space Administration and the National Science Foundation.
\end{acknowledgements}

\bibliographystyle{aa}
\bibliography{pm.bib}

\begin{appendix}

\section{Population Boxes in the VISTA near-infrared CMD}

Table \ref{spop} indicates the exact boundaries of the regions used in this study to select stars of a different type from the near-infrared CMD of the combined VMC-2MASS sample. The first column indicates the CMD region and Cols. 2, 3 and 4 mark the boundaries of each region, where values refer to 2MASS magnitudes, while Col. 5 lists the types of stars included in each region. The percentage of MW stars is shown within parenthesis.
CMD regions have been derived from \citet{2000ApJ...542..804N} using 2MASS All-Sky data and their Table $2$ values have been revised using their Figures and extended to include faint magnitudes from the 2MASS $6\times$ catalogue. In particular, region D has been split into two parts to distinguish MW stars from LMC stars.  

Similarly, Table \ref{vpop} indicates the exact boundaries to select stars of a different mean age from the near-infrared CMD of VMC stars reaching sensitivities well below the sensitivity of 2MASS observations. The first column indicates the CMD region and Cols. 2 and 3 mark the boundaries of each region, where values are in the Vega magnitude scale. Column 4 lists the type of stars and the median age of LMC stars while Col. 5 show the percentage of MW stars. CMD regions have been derived from the analysis of the star formation history within the tile LMC $8\_8$ by \citet{2012A&A...537A.106R}. In particular, this tile represents on average $A_V\sim 0.2$ mag and $(m-M)_0\sim18.39$ mag and contains old stars (log(t/yr)$>10.0$) and stars resulting from three further episodes of star formation. One at log(t/yr)$=9.9$ that formed $\sim31\%$ of the stellar mass, one at log(t/yr)$=9.1-9.3$ which formed $\sim21\%$ of the stellar mass and another at log(t/yr)$=8.5-8.7$ for stars in the part of the tile that is closer to the LMC centre with very little star formation at log(t/yr)$<8.3$. 

\begin{table*}
\caption{Stellar types in 2MASS data ($K_\mathrm{s}<15.5$ mag).}
\footnotesize
\label{spop}
\[
\begin{array}{cllll}
\hline \hline
\noalign{\smallskip}
\mathrm{Region} & \multicolumn{3}{c}{\mathrm{Boundaries}}  & \mathrm{Population}\\
\hline
\noalign{\smallskip}
\mathrm{A} & 11<K_\mathrm{s}<13.5  & K_\mathrm{s}>50\times(J-K_\mathrm{s})+3.5 &  & \mathrm{LMC\, supergiants\, and\, O\, dwarfs}\\
                     & K_\mathrm{s}>13.5         & (J-K_\mathrm{s})<0.2                                           &  & \\
\mathrm{B} & 5.5<K_\mathrm{s}<13.5 & (J-K_\mathrm{s})<0.5                                           & K_\mathrm{s}<50\times(J-K_\mathrm{s})+3.5    & \mathrm{MW\, F-K\, dwarfs\,(80\%),} \\
                     &                                            &                                                                                                                                    &                                                                                          & \mathrm{LMC\,supergiants}\\
\mathrm{C} & 5<K_\mathrm{s}<11       & (J-K_\mathrm{s})>0.5                                           & K_\mathrm{s}<-24\times(J-K_\mathrm{s})+32.6 & \mathrm{MW\, K\, dwarfs\,(80\%),} \\
                     & 11<K_\mathrm{s}<13.5  & 0.5<(J-K_\mathrm{s})<0.75                                & & \mathrm{LMC\, supergiants} \\
\mathrm{D}_\mathrm{L}  & K_\mathrm{s}>13.5 & K_\mathrm{s}>-10\times(J-K_\mathrm{s})+21 & K_\mathrm{s}< -15.7\times(J-K_\mathrm{s})+32.3 & \mathrm{LMC\,RGB,\,Early-AGB\, (95\%)}\\ 
\mathrm{D}_\mathrm{G} & K_\mathrm{s}>13.5 & (J-K_\mathrm{s})>0.2                                            & K_\mathrm{s}<-10\times(J-K_\mathrm{s})+21        & \mathrm{MW\,G-K\,dwarfs\,(75\%)}\\
\mathrm{E} & 12<K_\mathrm{s}<13.5  & K_\mathrm{s}>-10\times(J-K_\mathrm{s})+21                 & K_\mathrm{s}< -15.7\times(J-K_\mathrm{s})+32.3 & \mathrm{LMC\,RGB,\,tip\,of\,RGB\,stars}\\
\mathrm{F} & K_\mathrm{s}<12             & K_\mathrm{s}>-1.25\times(J-K_\mathrm{s})+12             & K_\mathrm{s}>-10\times(J-K_\mathrm{s})+21         & \mathrm{LMC\,O-rich\,AGB\,stars}\\
                     &                                             & K_\mathrm{s}< -15.7\times(J-K_\mathrm{s})+32.3         & & \\
\mathrm{G} & K_\mathrm{s}<-1.25\times(J-K_\mathrm{s})+12       &  K_\mathrm{s}>-10\times(J-K_\mathrm{s})+21 & K_\mathrm{s}< -15.7\times(J-K_\mathrm{s})+32.3 & \mathrm{LMC\,massive\,AGB\,stars}\\
\mathrm{H} & 7<K_\mathrm{s}<11 & K_\mathrm{s}>-24\times(J-K_\mathrm{s})+32.6                                                     & K_\mathrm{s}<-10\times(J-K_\mathrm{s})+21        & \mathrm{LMC\,K-M\,supergiants,} \\
                      &                                     &                                                                                                                                          &                                                                                          & \mathrm{MW\,M\,dwarfs\,and\,K-M\,giants}\\
\mathrm{I}   & 11<K_\mathrm{s}<13.5 & (J-K_\mathrm{s})>0.75                                                                                          & K_\mathrm{s}<-10\times(J-K_\mathrm{s})+21  & \mathrm{LMC\,supergiants,} \\
                     &                                            &                                                                                                                                    &                                                                                          & \mathrm{MW\,K-M\,dwarfs\,(55\%)}\\
\mathrm{J}  & (J-K_\mathrm{s})<2        & K_\mathrm{s}>-1.4\times(J-K_\mathrm{s})+11.8                                              & K_\mathrm{s}<-1.4\times(J-K_\mathrm{s})+13.8  & \mathrm{LMC\,C\,stars}\\
                     &                                             & K_\mathrm{s}>-15.7\times(J-K_\mathrm{s})+32.3                                           & & \\
\mathrm{K} & 2<(J-K_\mathrm{s})<5   & K_\mathrm{s}>0.67\times(J-K_\mathrm{s})+6.6                                                & K_\mathrm{s}<0.67\times(J-K_\mathrm{s})+10.4 & \mathrm{LMC\,dusty\,AGB\,stars}\\
\mathrm{L} & (J-K_\mathrm{s})<2.5     & K_\mathrm{s}>-15.7\times(J-K_\mathrm{s})+32.3                                            &  K_\mathrm{s}>0.6\times(J-K_\mathrm{s})+11.25 &  \mathrm{LMC\,red\,RGB,} \\
                     &                                            &                                                                                                                                    &                                                                                          & \mathrm{MW\,M\,and\,L\,dwarfs,\,galaxies}\\
\noalign{\smallskip}
\hline
\end{array}
\]
\end{table*}

\begin{table*}
\caption{Stellar ages in VMC data ($K_\mathrm{s}>13$ mag).}
\footnotesize
\label{vpop}
\[
\begin{array}{clllcc}
\hline \hline
\mathrm{Region} & \multicolumn{2}{c}{\mathrm{Boundaries}}  & \mathrm{log(Age)} & \mathrm{MW}\\
 & & & (\mathrm{yr}) & \% \\
\hline
\noalign{\smallskip}
\mathrm{A} & K_\mathrm{s}<19                                                          & (J-K_\mathrm{s})>-0.2                                                  & \mathrm{LMC}\,7.95\pm0.27  & 0 \\
                     & K_\mathrm{s}>7.5\times(J-K_\mathrm{s})+16.5      & K_\mathrm{s}<-25\times(J-K_\mathrm{s})+16.5     & \\
\mathrm{B} & K_\mathrm{s}<19                                                          & K_\mathrm{s}>-25\times(J-K_\mathrm{s})+16.5      &  \mathrm{LMC}\,8.51\pm0.16 & 0 \\
                     & K_\mathrm{s}>3.333\times(J-K_\mathrm{s})+16.5 & K_\mathrm{s}<-13.333\times(J-K_\mathrm{s})+19 & \\
\mathrm{C} & K_\mathrm{s}<19                                                          & K_\mathrm{s}>-13.333\times(J-K_\mathrm{s})+19 & \mathrm{LMC}\,8.87\pm0.14 &1 \\
                     & K_\mathrm{s}>2\times(J-K_\mathrm{s})+16.7         & K_\mathrm{s}<-6\times(J-K_\mathrm{s})+19.9         & \\
\mathrm{D} & K_\mathrm{s}<19                                                         & K_\mathrm{s}>-6\times(J-K_\mathrm{s})+19.9          &  \mathrm{LMC}\,9.19\pm0.14 & 2 \\
                     & K_\mathrm{s}>-2.5\times(J-K_\mathrm{s})+18.5    & K_\mathrm{s}<-15\times(J-K_\mathrm{s})+25           & \\
\mathrm{E} & 16<K_\mathrm{s}<19                                                  & K_\mathrm{s}<-30\times(J-K_\mathrm{s})+37            &  \mathrm{LMC}\,9.53\pm0.36 & 0.3 \\
                     &                                                                                         & K_\mathrm{s}>-15\times(J-K_\mathrm{s})+25             & \\
\mathrm{F} & K_\mathrm{s}<19                                                         & (J-K_\mathrm{s})<0.9                                                       &  \mathrm{LMC}\,9.81\pm0.19 & 94\\
                     & K_\mathrm{s}>-30\times(J-K_\mathrm{s})+37       &  K_\mathrm{s}>-8.75\times(J-K_\mathrm{s})+22.125 & \\
\mathrm{G} & 12.5<K_\mathrm{s}<16                                              & 0<(J-K_\mathrm{s})<0.2                                                  & \mathrm{LMC}\,8.13\pm0.10 & 13 \\
\mathrm{H} & 12.5<K_\mathrm{s}<16                                              & (J-K_\mathrm{s})>0.2                                                      & \mathrm{LMC}\,9.75\pm0.76 & 77\\
                      &                                                                                        & K_\mathrm{s}<-15\times(J-K_\mathrm{s})+22            & \\
\mathrm{I} & K_\mathrm{s}>12.5 & (J-K_\mathrm{s})>0.4 &  \mathrm{LMC}\,8.61\pm0.17 & 15\\
 & K_\mathrm{s}>-15\times(J-K_\mathrm{s})+22 &  K_\mathrm{s}<-9.167\times(J-K_\mathrm{s})+20.917 & \\
 & K_\mathrm{s}<-15\times(J-K_\mathrm{s})+25 &  & \\
\mathrm{J} & (J-K_\mathrm{s})>0.4 & K_\mathrm{s}>-9.167\times(J-K_\mathrm{s})+20.917 &  \mathrm{LMC}\,9.15\pm0.60 & 1 \\
 & K_\mathrm{s}<-2.5\times(J-K_\mathrm{s})+18.5 & K_\mathrm{s}<-15\times(J-K_\mathrm{s})+25 &  \\
\mathrm{K} & 12.5<K_\mathrm{s}<16 & K_\mathrm{s}>-15\times(J-K_\mathrm{s})+25 &  \mathrm{LMC}\,9.45\pm0.37 & 3 \\
 & & K_\mathrm{s}<-8.75\times(J-K_\mathrm{s})+22.125 & \\
\mathrm{L} & 12.5<K_\mathrm{s}<19 & 0.9<(J-K_\mathrm{s})<2.2 & \mathrm{Galaxies} & 3 \\
 & & K_\mathrm{s}>-8.75\times(J-K_\mathrm{s})+22.125 & \\
\noalign{\smallskip}
\hline
\end{array}
\]
\end{table*}

\section{Proper motion trends}
\begin{figure*}
\resizebox{\hsize}{!}{\includegraphics{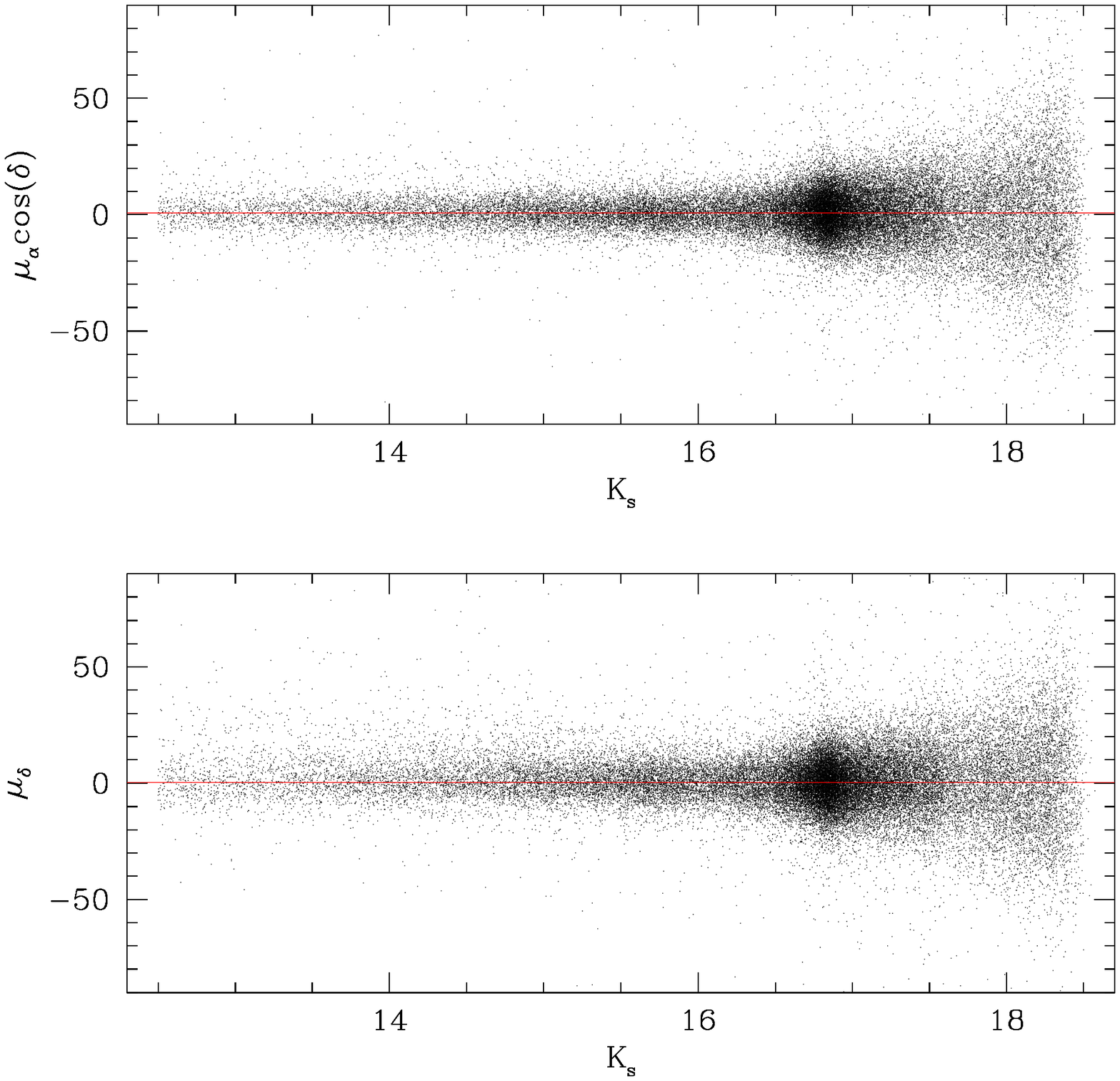}
\includegraphics{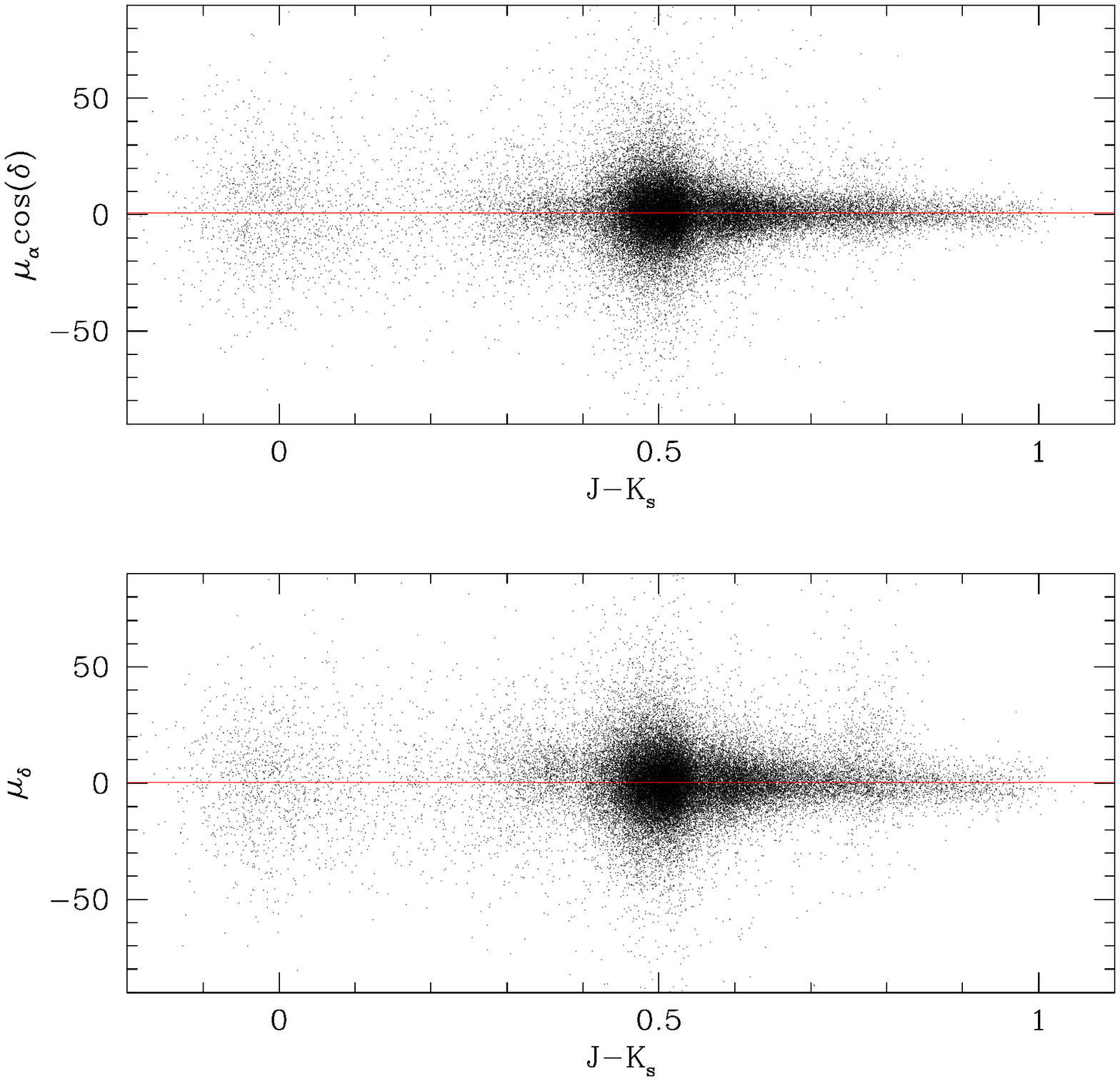}}
\caption{Proper motion in mas yr$^{-1}$ as function of $K_\mathrm{s}$ magnitude (left) and $(J-K_\mathrm{s})$ colour (right) for all LMC sources of Fig. \ref{vmcbox}.}
\label{tmagcollmc}
\end{figure*}

\begin{figure*}
\resizebox{\hsize}{!}{\includegraphics{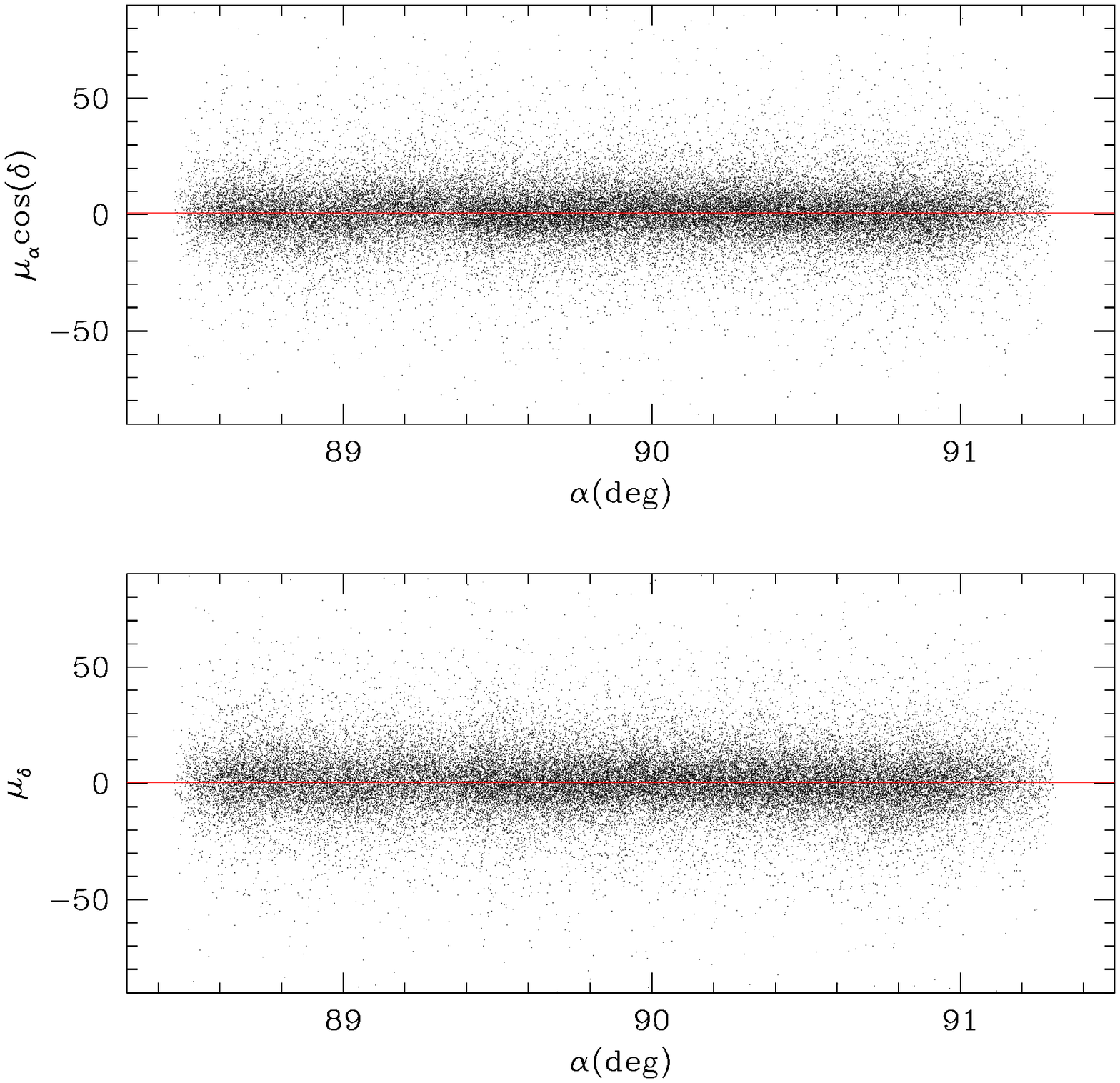}
\includegraphics{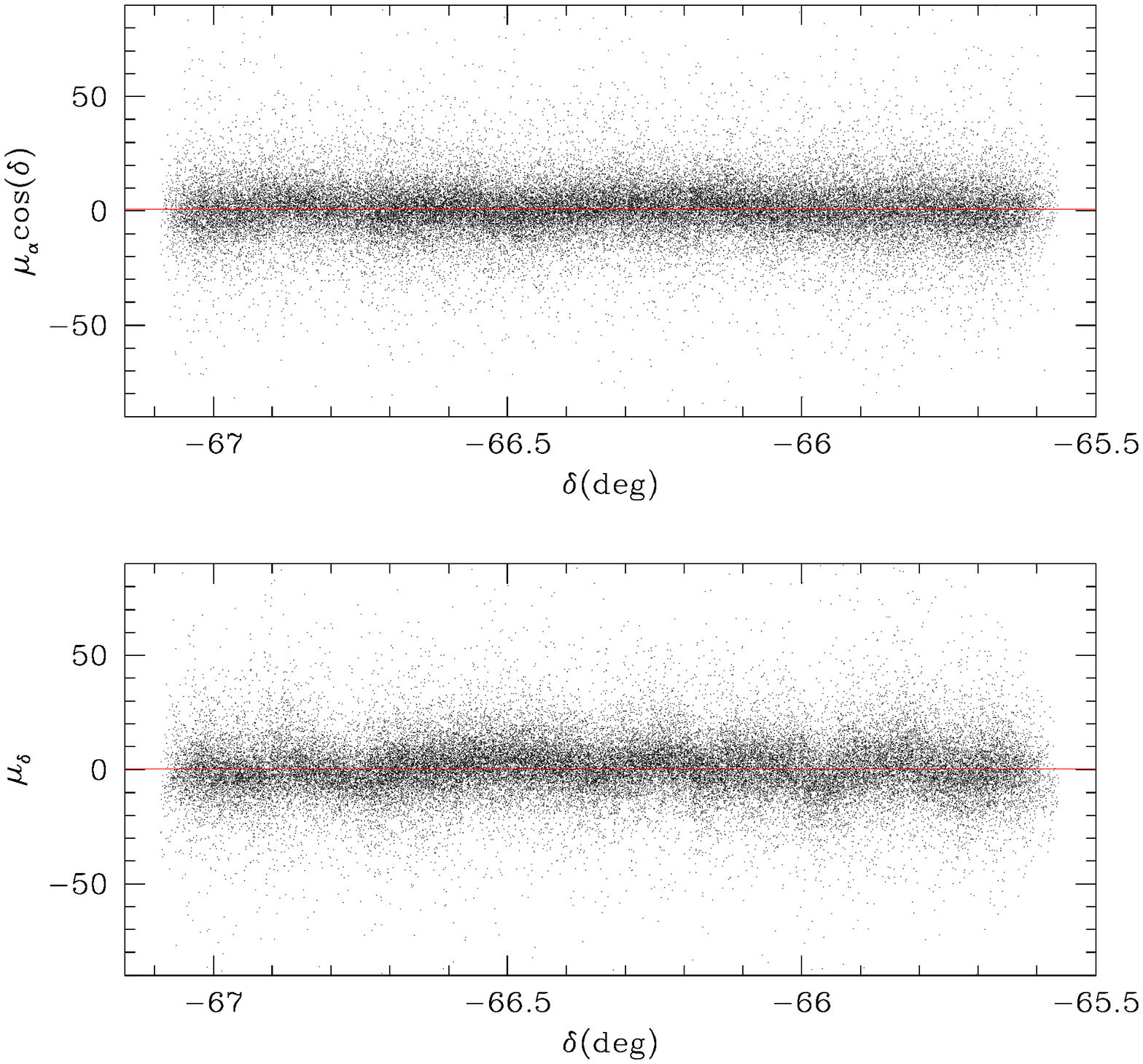}}
\caption{Proper motion in mas yr$^{-1}$ as function of right ascension (left) and declination (right) for all LMC sources of Fig. \ref{vmcbox}.}
\label{tradeclmc}
\end{figure*}

\begin{figure*}
\resizebox{\hsize}{!}{\includegraphics{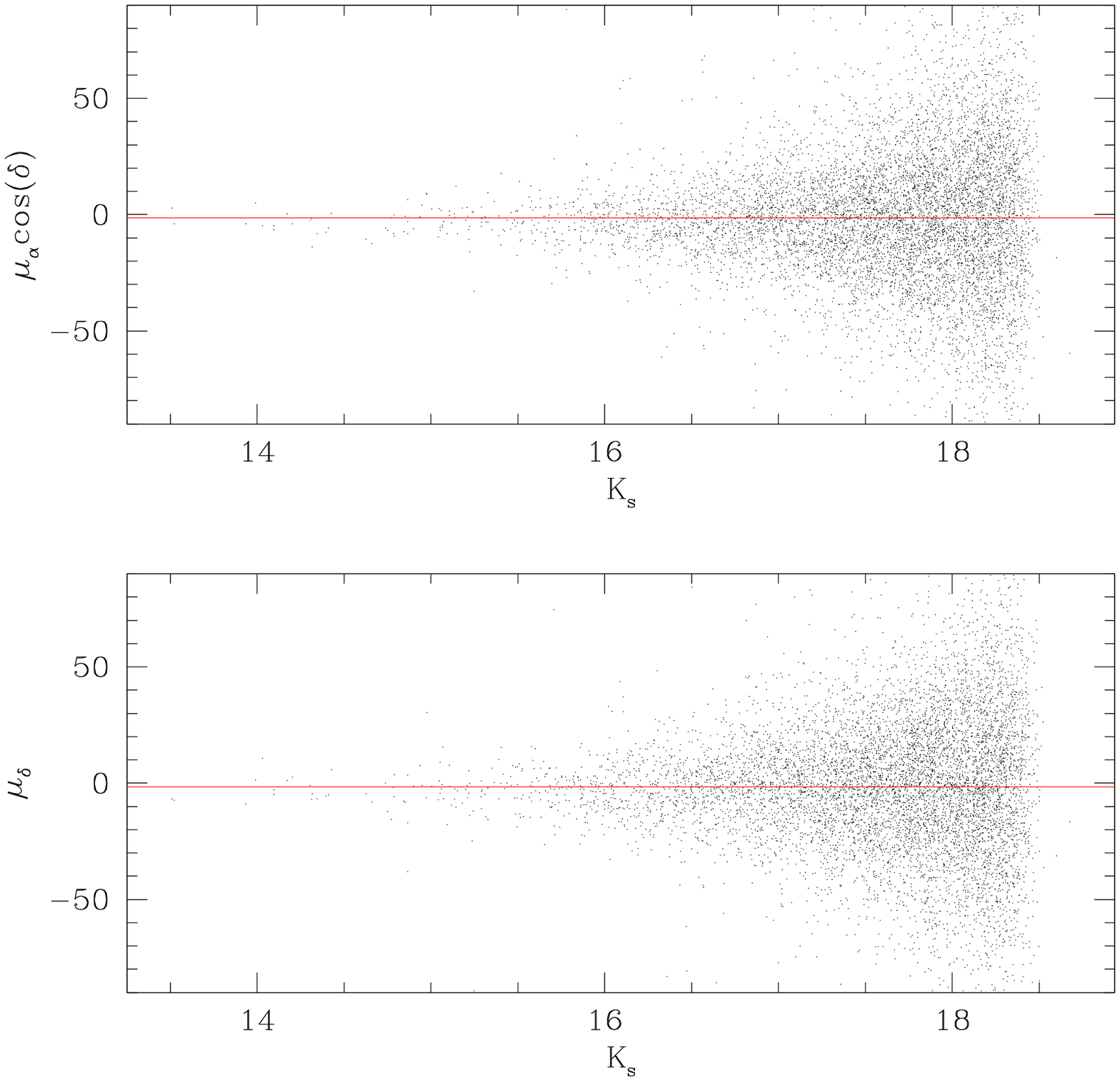}
\includegraphics{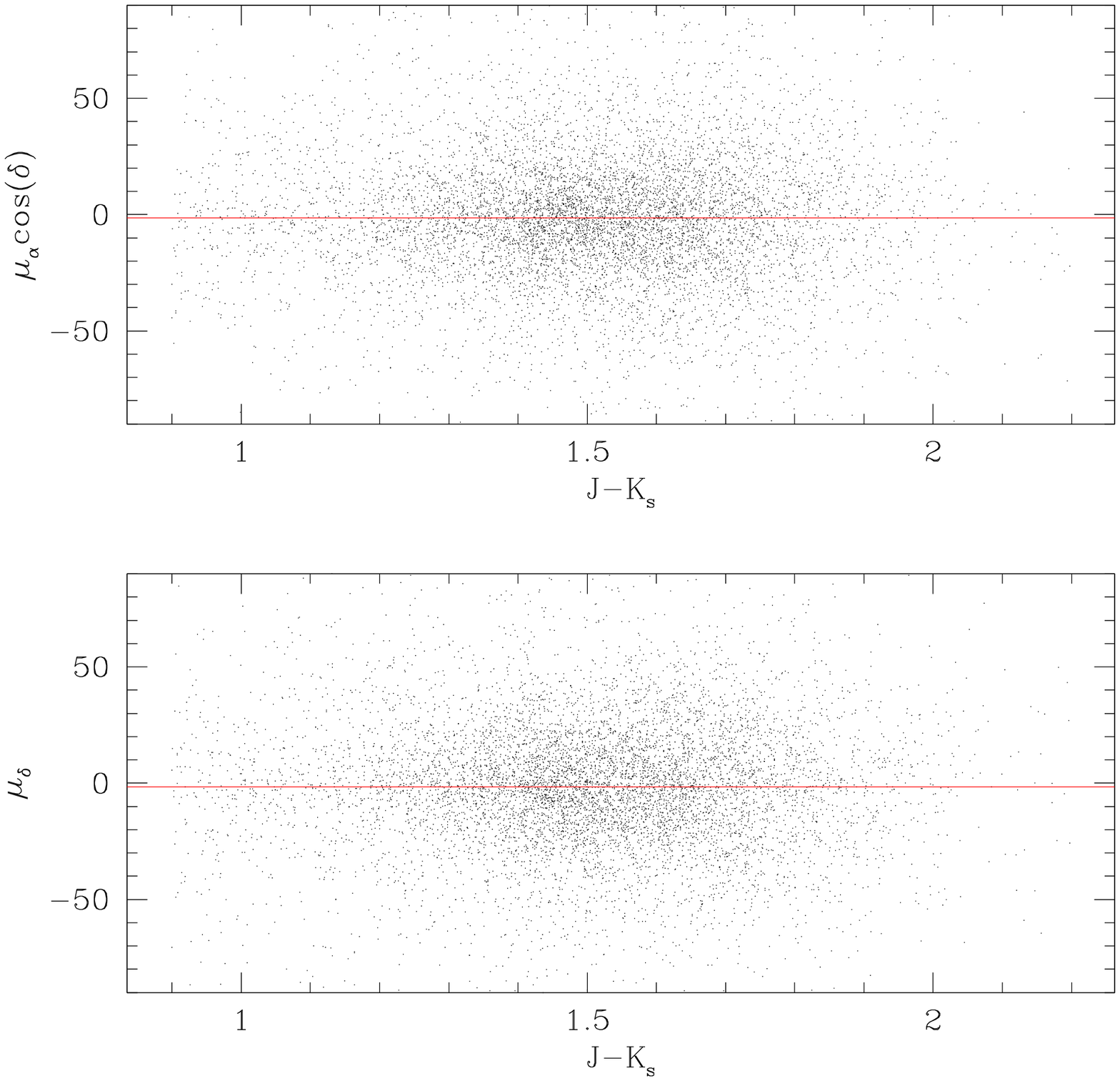}}
\caption{Proper motion in mas yr$^{-1}$ as function of $K_\mathrm{s}$ magnitude (left) and $(J-K_\mathrm{s})$ colour (right) for galaxies in region L of Fig. \ref{vmcbox}.}
\label{tmagcolgal}
\end{figure*}

\begin{figure*}
\resizebox{\hsize}{!}{\includegraphics{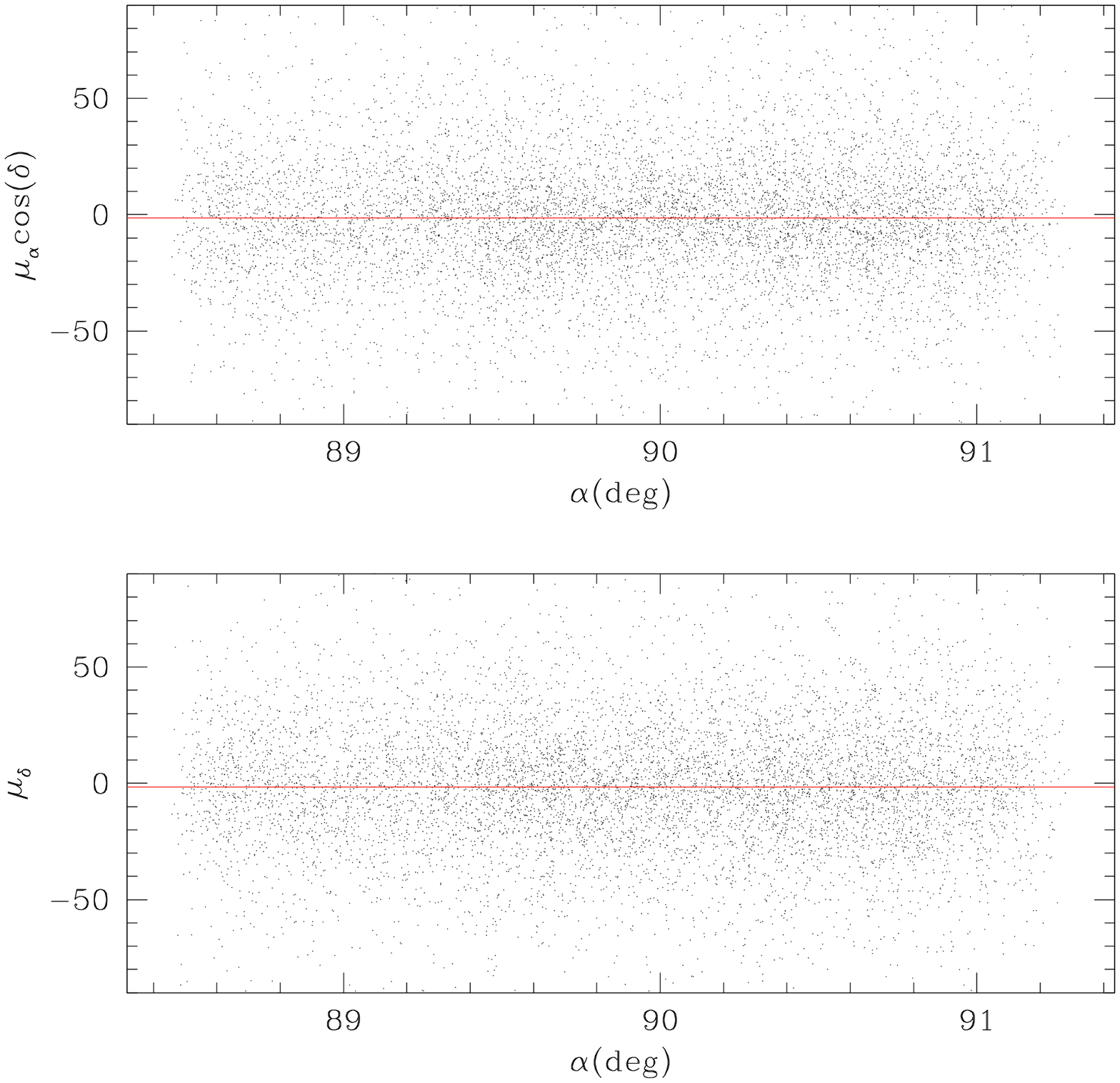}
\includegraphics{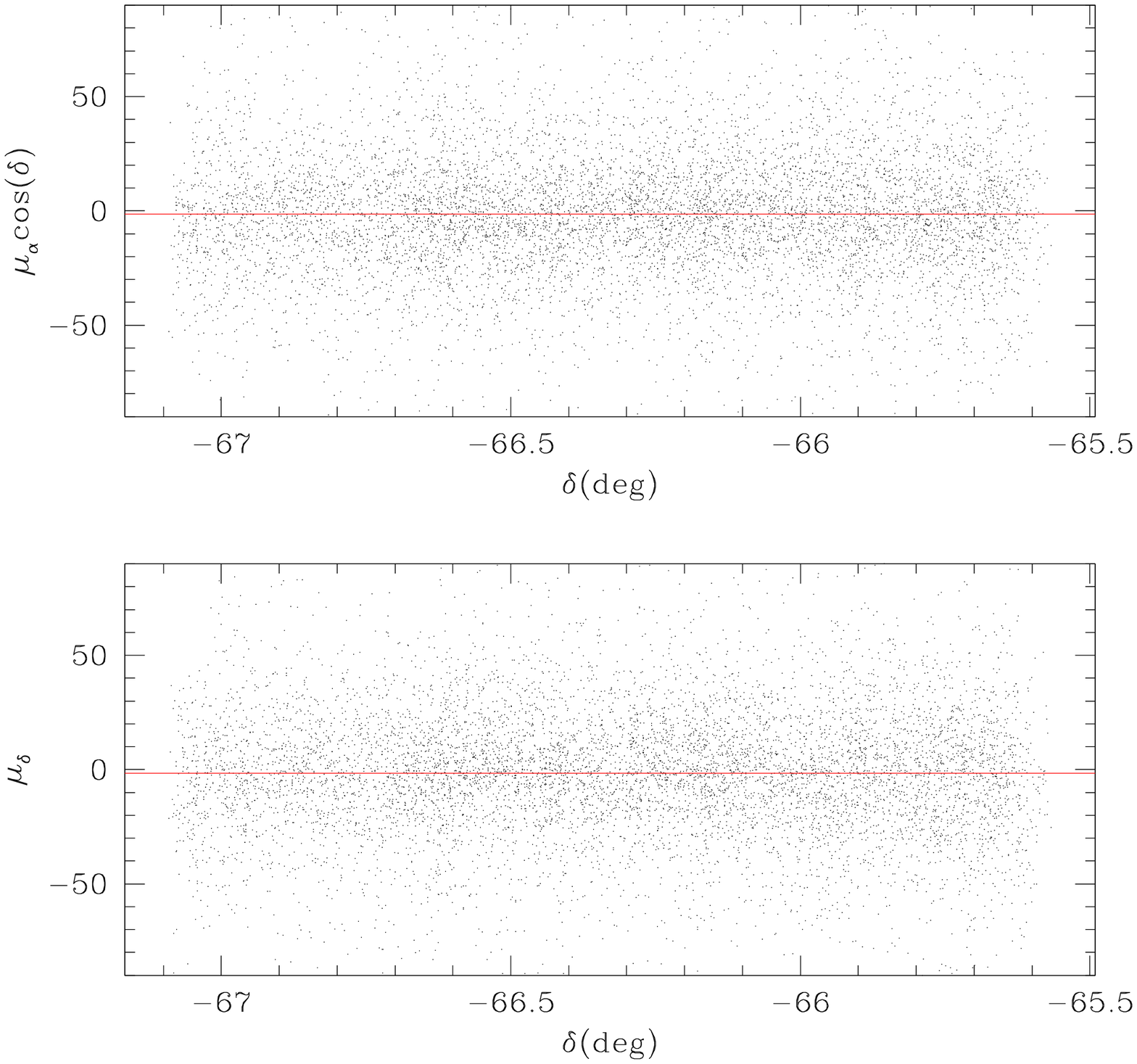}}
\caption{Proper motion in mas yr$^{-1}$ as function of right ascension (left) and declination (right) for galaxies in region L of Fig. \ref{vmcbox}.}
\label{tradecgal}
\end{figure*}

Figures \ref{tmagcollmc} and \ref{tradeclmc} show the PM as a function of $K_\mathrm{s}$ magnitude, $(J-K_\mathrm{s})$ colour and position, right ascension ($\alpha$) and declination ($\delta$) for LMC stars in regions $ABCDEGHIJK$ of  Fig. \ref{vmcbox}. Similar plots for galaxies in region $L$ are shown in Figs. \ref{tmagcolgal} and \ref{tradecgal}. Horizontal lines through the data indicate the corresponding mean PM for stars in the region.  An oscillating component, clearly visible in the bottom-right panel of Fig. \ref{tradeclmc} ($\mu_\delta$ versus $\delta$), appears symmetric with respect to the centre of the distribution and would not therefore influence the resulting mean PMs. This effect is probably due to astrometric differences among the VISTA detectors where $8$ of them approximately cover each axis. Note that these figures include all points prior to the $3\sigma$ clipping procedure applied to derive mean PM values.

\end{appendix}

\end{document}